\documentstyle[12pt]{article}

\font\blackboard=msbm10 at 12pt
\font\blackboards=msbm7
\font\blackboardss=msbm5
\newfam\black
\textfont\black=\blackboard
\scriptfont\black=\blackboards
\scriptscriptfont\black=\blackboardss

\newcommand{\junk}[1]{}

\newcommand{\ba}{\begin{array}}
\newcommand{\ea}{\end{array}}
\newcommand{\be}{\begin{equation}}
\newcommand{\ee}{\end{equation}}
\newcommand{\bea}{\begin{eqnarray}}
\newcommand{\eea}{\end{eqnarray}}
\newcommand{\beas}{\begin{eqnarray*}}
\newcommand{\eeas}{\end{eqnarray*}}

\def\laplace{{\kern1pt\vbox{\hrule height 1.2pt\hbox{\vrule width
1.2pt\hskip
  3pt\vbox{\vskip 6pt}\hskip 3pt\vrule width 0.6pt}\hrule height
  0.6pt}
  \kern1pt}}
\def\scriptlap{{\kern1pt\vbox{\hrule height 0.8pt\hbox{\vrule width
  0.8pt
  \hskip2pt\vbox{\vskip 4pt}\hskip 2pt\vrule width 0.4pt}\hrule height
  0.4pt}
  \kern1pt}}
\def\slash#1{{\rlap{$#1$} \thinspace /}}
\def\roughly#1{\raise.3ex\hbox{$#1$\kern-.75em\lower1ex\hbox{$\sim$}}}

\newcommand{\NP}{{\em Nucl.\ Phys.\ }}

\newcommand{\PR}{{\em Phys.\ Rev.\ }}

\newcommand{\gone}[1]{}
\begin{document}
\pagestyle{plain}
\setcounter{page}{1}

\baselineskip16pt

\begin{titlepage}

\begin{flushright}

\end{flushright}
\vspace{8 mm}

\begin{center}

{\Large \bf Hopf-Wess-Zumino term in the effective action of the 6d, (2, 0) field theory revisted\\}
%\vspace{3mm}

\end{center}

\vspace{7 mm}

\begin{center}

{\bf Shan Hu $^{1,a}$, and Dimitri Nanopoulos $^{1,2,3,b}$}

\vspace{3mm}
{\small \sl $^{1}$George P. and Cynthia W.Mitchell Institute for Fundamental Physics, Texas A\&M University,} \\
{\small \sl  College Station, TX 77843, USA} \\
{\small \sl $^{2}$Astroparticle physics Group, Houston Advanced Research Center (HARC),} \\

{\small \sl Mitchell Campus, Woodlands, TX 77381, USA} \\

{\small \sl $^{3}$Academy of Athens, Division of Nature Sciences,} \\

{\small \sl 28 panepistimiou Avenue, Athens 10679, Greece} \\

\vspace{3mm}

{\small \tt $^{a}$hushan@physics.tamu.edu, $^{b}$dimitri@physics.tamu.edu} \\

\end{center}

\vspace{8 mm}

\begin{abstract}

We discuss the Hopf-Wess-Zumino term in the effective action of the $6d$ $(2, 0)$ theory of the type $A_{N-1}$ in a generic Coulomb branch. For such terms, the supergravity calculation could be trusted. We calculate the WZ term on supergravity side and show that it could compensate the anomaly deficit, as is required by the anomaly matching condition. In contrast with the SYM theory, in which each WZ term involves one root $e_{i}-e_{j}$, here, the typical WZ term involves two roots $e_{i}-e_{j}$ and $e_{k}-e_{j}$. Such kind of triple interaction may come from the integrating out of the massive states carrying three indices. A natural candidate is the recently proposed $1/4$ BPS objects in the Coulomb phase of the $6d$ $(2, 0)$ theories. The WZ term could be derived from the field theory by the integration out of massive degrees of freedom. Without the $6d$ $(2, 0)$ theory at hand, we take the supersymmetric equations for the 3-algebra valued (2, 0) tensor multiplet as the prototype to see how far we can go. The $H_{3}\wedge A_{3}$ part of the WZ term is obtained, while the $A_{3}\wedge F_{4}$ part, which is the term accounting for the anomaly matching, cannot be produced by the standard fermion loop integration.

\end{abstract}

%\vspace{2cm}
\vspace{1cm}
\begin{flushleft}


\end{flushleft}
\end{titlepage}
\newpage

\section{Introduction}
Low energy effective action of the field theory in the Coulomb branch may contain the Wess-Zumino term arising from the integration out of massive fermions getting masses via the Yukawa coupling with the vacuum expectation value of the scalar fields \cite{1, 2}. The existence of the Wess-Zumino term is also required by the anomaly matching condition \cite{3}. At a generic point of the moduli space, the gauge symmetry is broken, and then, the 't Hooft anomaly produced by massless degrees of freedom is different from the anomaly at the origin. On the other hand, the anomaly matching condition states that the 't Hooft anomaly should be the same everywhere on the moduli space of vacua. As a result, away from the origin, the integrating out of the massive degrees of freedom should generate the Wess-Zumino term in the low energy effective action compensating the deficit so that the total anomaly remains the same \cite{3, A}.

The WZ term is a topological term that does not depend on the metric nor the coupling, so it is protected without the need of invoking any supersymmetric non-renormalization theorems. For such terms, we may expect that the 1-loop calculation in field theory and the supergravity calculation would match. On supergravity side, the Wess-Zumino term is associated with the magnetic-electric coupling. For Dp-branes with $p \geq 3$, it is given by $\int_{W_{p+2}} F_{8-p}(\wedge dA)^{p-3} = (-1)^{p}\int_{W_{p+1}} F_{8-p}\wedge A (\wedge dA)^{p-4}$ \cite{D, E}, where $F_{8-p}$ is the magnetic field strength, while $(\wedge dA)^{p-3}$ offers the electric charge. When $p=3$, $\int_{W_{p+2}} F_{8-p}(\wedge dA)^{p-3} \rightarrow \int_{W_{5}} F_{5}$, because the D3-brane carries magnetic as well as the electric charge \cite{E}. For M5 branes, the WZ term is composed by $\int_{W_{6}}db_{2}\wedge A_{3}$ and $\int_{W_{7}}A_{3} \wedge F_{4}$, which are discussed in \cite{J} and \cite{A} respectively. $\int_{W_{6}}db_{2}\wedge A_{3}$ does not contribute to the anomaly. It is $\int_{W_{7}}A_{3} \wedge F_{4}$ that accounts for the anomaly deficit. \cite{A} considered the situation when the gauge symmetry is broken from $SU(N+1)$ to $SU(N)\times U(1)$ by the vacuum expectation value $\phi^{a}$. The corresponding WZ term takes the form of $\int_{W_{7}}\sigma_{3}(\hat{\phi}) \wedge d \sigma_{3}(\hat{\phi})$, where $d \sigma_{3}(\hat{\phi})$ is the pullback of the 4-form field strength generated by a single M5 brane while $\sigma_{3}(\hat{\phi}) $ is the corresponding 3-form potential. $\hat{\phi} = \phi/|\phi|$. It was shown that with the coefficient given by $N(N+1)/2$, the WZ term could reproduce the anomaly deficit between $SU(N+1)$ and $SU(N)\times U(1)$. In this note, we will extend the discussion to the generic Coulomb branch $(\phi^{a}_{1},\cdots,\phi^{a}_{N})$. We will show that the supergravity calculation could give the right coefficient, while the WZ term, although takes the form of $\int_{W_{7}}\sigma_{3}(\hat{\phi}_{ij}) \wedge d \sigma_{3}(\hat{\phi}_{kj})$ with $\hat{\phi}_{ij} = (\phi_{i}-\phi_{j})/|\phi_{i}-\phi_{j}|$, could produce the same amount of anomaly as that of $\int_{W_{7}}\sigma_{3}(\hat{\phi}) \wedge d \sigma_{3}(\hat{\phi})$. So the WZ term obtained from the supergravity calculation indeed compensates the anomaly deficit thus should appear in the low energy effective action, as is required by the anomaly matching condition.

In the generic Coulomb branch, the WZ term in SYM theory is $(-1)^{p}\int_{W_{p+1}} F_{8-p}(\hat{\phi}_{ij})\wedge A_{ij}[\wedge dA_{ij}]^{p-4}$,\footnote{$A_{ij} = A_{i} - A_{j}$. It is the relative flux that makes sense.} which is the typical pair-wise interaction arising from the the integration out of massive fermions carrying index $(i, j)$, or open strings connecting the $i_{th}$ and the $j_{th}$ D-brane \cite{D, E}. The term $\int_{W_{7}}\sigma_{3}(\hat{\phi}_{ij}) \wedge d \sigma_{3}(\hat{\phi}_{kj})$ for M5 branes seems indicate some kind of triple interaction: three M5 branes could interact simultaneously. One may naturally expect that such term comes from the integration out of massive fermions with $(i, j, k)$ index, or open M2 branes connecting the $i_{th}$, the $j_{th}$, and the $k_{th}$ M5 branes. In \cite{K} and more recently, \cite{L}, the $1/4$ BPS objects in the Coulomb phase of the ADE-type 6d $(2, 0) $ superconformal theories are considered. They are made of waves on selfdual strings and junctions of selfdual strings. In \cite{L}, it was shown that the number of $1/4$ BPS objects matches exactly one third of the anomaly constant $c_{G} = d_{G}h_{G}$ for all ADE types, indicating that the anomaly may be produced by these $1/4$ BPS objects. Moreover, the tension of the string junctions is characterized by $(|\phi_{i}- \phi_{j}|, |\phi_{j}- \phi_{k}|, |\phi_{k}- \phi_{i}|)$, which is just what is needed to produce the WZ term, since the selfdual string with tension $|\phi_{i}- \phi_{j}|$ is not enough to give the coupling like $\sigma_{3}(\hat{\phi}_{ij}) \wedge d \sigma_{3}(\hat{\phi}_{kj})$.

For SYM theories, the WZ term could be derived by a 1-loop calculation \cite{D, E, F}. It is expected that the WZ term for $6d$ $(2, 0)$ theories could also be obtained via a proper integration. The $6d$ $(2, 0)$ theory is not constructed yet. Nevertheless, the form of the WZ term may offer some hint on the possible structure the underlying theory. Of course, it is possible that to construct the $6d$ SCFT, some new ingredient must be added making the theory different from the QFT in normal sense, and so the way to calculate the WZ term is also beyond the present knowledge. On the other hand, if the $6d$ $(2, 0)$ theory could be built as an ordinary quantum field theory just as that for M2 branes, we will be able to calculate the WZ term with the standard field theory methods. In \cite{Z}, the supersymmetric equations of motion for the 3-algebra valued $(2,0)$ tensor multiplet were found, which may shed light on our understanding of the mysterious $6d$ SCFT. We will calculate the WZ term for the 3-algebra valued $(2,0)$ tensor multiplet to see how far we can go. The $H_{3}\wedge A_{3}$ part of the WZ term is obtained. Especially, without the constraint equations in \cite{Z}, $H_{3}\wedge A_{3}$ cannot be derived. However, the $A_{3}\wedge F_{4}$ part cannot be obtained by the 1-loop fermion integration, so either a refined calculation method or a refined theory is needed.

This paper is organized as follows: In section 2, we get the WZ term for $6d$ $A_{N-1}$ $(2, 0)$ theory in a generic Coulomb branch from the supergravity calculation. Part of the details is given in appendix A. In section 3, we show that the WZ terms obtained from the supergravity calculation could indeed compensate the anomaly deficit thus guarantee the anomaly matching condition. In section 4, we discuss the possible degrees of freedom in M5 branes producing the WZ term. In section 5, we caculate the WZ term for the 3-algebra valued $(2, 0)$ tensor multiplet. The conclusion is in section 6.

\section{The Hopf-Wess-Zumino term from the supergravity calculation}
Consider the $6d$ $(2, 0)$ field theory describing $N$ M5 branes. On a generic Coulomb branch $(\phi^{a}_{1},\cdots,\phi^{a}_{N})$ with $a = 1\cdots5$ and $\phi_{i}\neq \phi_{j}$, for $i\neq j$, the gauge symmetry is broken to $U(1)^{N}$. On supergravity side, we have $N$ M5 branes locating at $(\phi_{1},\cdots,\phi_{N})$. On field theory side, $N$ copies of $(2, 0) $ tensor multiplets remain massless, while the rest fields get masses. Integrating out these massive degrees of freedom, one may obtain the effective action of the $6d$, $(2, 0)$ field theory on Coulomb branch. At least for WZ terms, the calculation on both sides should coincide. $6d$ $(2, 0) $ field theory is still mysterious to us, while the multi-centered supergravity solution of M5 branes is more tractable, so we will try to get the WZ term in the effective action through the supergravity calculation. 

The action for the coupling of M5 branes with the 11d supergravity could be written as\footnote{Just as the (6.15) in \cite{a2}, $*F_{4}$ could be added into the worldvolume of the M5 brane, but then an equal term will appear in the bulk.} \cite{a1,a11,a2}
\begin{eqnarray} \label{a}
S &=&  S_{g}+S_{M5}
 \nonumber \\
&=& \frac{1}{2 \kappa^{2}}\int_{M_{11}} *R-\frac{1}{2} *\hat{F}_{4}\wedge \hat{F}_{4} - \frac{1}{6}F_{4}\wedge F_{4} \wedge A_{3} \nonumber \\
&-& T_{5}\int_{W_{6}} d^{6}\xi \sqrt{-\det(g_{\mu \nu}+ (i_{v_{1}} \tilde{*} h_{3})_{\mu \nu})} + \frac{1}{2} v_{1}\wedge h_{3} \wedge \tilde{*} (v_{1}\wedge \tilde{*} h_{3})\nonumber \\
&+& \frac{T_{5}}{2} \int_{W_{6}} db_{2} \wedge A_{3} + \frac{T_{5}}{2} \int_{W_{7}} A_{3} \wedge F_{4}
\end{eqnarray} 
where $F_{4} = dA_{3}$, $W_{6} = \partial W_{7}$,  
\begin{equation}\label{h}
	\hat{F}_{4} = F_{4} + 2 \kappa^{2} T_{5} * G_{7},
\end{equation}
\begin{equation}
	h_{3} = db_{2}-A_{3}.
\end{equation}
$d* G_{7} = *J_{6}$. $*J_{6}$ is the the M5-brane current. The last term in (\ref{a}) is just the Hopf-Wess-Zumino term proposed in \cite{A}. The field equations for $\hat{F}_{4}$ are 
\begin{equation}\label{101}
	d \hat{F}_{4} = 2\kappa^{2} T_{5} *J_{6}, 
\end{equation}
\begin{equation}\label{100}
	d * \hat{F}_{4} + \frac{1}{2}\hat{F}_{4}\wedge \hat{F}_{4} = - 2\kappa^{2} T_{5} \; h_{3}\wedge *J_{6}. 
\end{equation}
$* G_{7}$, $A_{3}$, and $F_{4}$ have the dependence on gauge, while $*J_{6}$, $h_{3}$, and $\hat{F}_{4}$ are gauge independent. (\ref{101}) and (\ref{100}) only contain gauge invariant quantities.

Suppose the vacuum expectation values of $b_{2}$ are equal to zero, consider $N$ M5 branes locating at $(\phi_{1},\cdots,\phi_{N})$. The WZ term is related with the electric-magnetic coupling, so we only need to calculate the magnetic field generated by M5 branes in the given configuration, which, in present case, is  
\begin{equation}
	\hat{F}_{4} = \sum^{N}_{i = 1} \hat{F}_{4i} = Q_{1}\sum^{N}_{i = 1} \omega_{4i}, 
\end{equation} 
where $\omega_{4i}$ is the unite volume form of $S^{4}$ surrounding the $i_{th}$ brane. The corresponding 3-form field is
\begin{equation}
	A_{3} = \sum^{N}_{i = 1} A_{3i}= Q_{1}\sum^{N}_{i = 1} \sigma_{3i}.  
\end{equation}
$d \sigma_{3i} = \omega_{4i}$. To calculate the WZ term on the $j_{th}$ M5 brane, we need the pullback of $A_{3}$ and $F_{4}$ on the corresponding $W_{7}$. The pullback of $* G_{7}$ on $W_{7}$ vanishes, so we simply have 
\begin{equation}
\int_{W_{7j}} A_{3} \wedge F_{4} = \int_{W_{7j}} A_{3} \wedge \hat{F}_{4}= Q_{1}^{2} \int_{W_{7j}}  \sum^{N}_{i = 1} \sigma_{3ij} \wedge \sum^{N}_{k = 1} \omega_{4kj}, 	
\end{equation}
where $d \sigma_{3ij} = \omega_{4ij}$. $\omega_{4ij}$ is the pullback of $\omega_{4i}$ on $W_{7j}$. Altogether, 
\begin{equation}\label{b}
	\frac{T_{5}}{2} \sum^{N}_{j = 1} \int_{W_{7j}} A_{3} \wedge F_{4} = \frac{Q_{1}^{2} T_{5}}{2}  \sum^{N}_{i = 1} \sum^{N}_{j = 1} \sum^{N}_{k = 1} \int_{W_{7j}}    \sigma_{3ij} \wedge \omega_{4kj}
\end{equation}
Aside from the $F_{4}\wedge F_{4}\wedge A_{3}$ term in supergravity, $\int A_{3}\wedge F_{4}$ is the other term which has the $N^{3}$ scaling.

However, (\ref{b}) is still not exactly the WZ term in the effective action. First, when $i=j=k$, we get a self-interaction term. There are totally $N$ such terms. These self-interaction terms will be produced only after the $N$ massless tensor multiplets are also integrated out. Since we only integrate massive degrees of freedom, these terms will not appear in the effective action. Second, as is shown in the Matrix theory calculation \cite{C}, for the given brane configuration, or equivalently, the Coulomb branch, the more accurate expression for the effective action on supergravity side should be 
\begin{equation}\label{c}
	S_{eff} = S_{g}+S_{M5}, 
\end{equation}
where $S_{g}$ is the action of the supergravity fields generated by M5 branes, while $S_{M5}$ is the action of M5 branes on the background generated by themselves. $S_{eff}$ is on-shell with respect to supergravity as it should be. (\ref{b}) comes from $S_{M5}$. In appendix A, we will show that $S_{g}$ contains a term which is $-2/3$ of (\ref{b}), so altogether, we have
\begin{equation}
\Gamma_{WZ} = \frac{Q_{1}^{2} T_{5}}{6} ( \sum^{N}_{i = 1} \sum^{N}_{j = 1} \sum^{N}_{k = 1}\int_{W_{7j}} \sigma_{3ij} \wedge \omega_{4kj}-\sum^{N}_{i=1}\int_{W_{7i}} \sigma_{3ii} \wedge \omega_{4ii})
\end{equation}
or 
\begin{equation}\label{j}
	2\kappa^{2} \Gamma_{WZ} = \frac{Q_{1}^{3} }{6} ( \sum^{N}_{i = 1} \sum^{N}_{j = 1} \sum^{N}_{k = 1}\int_{W_{7j}} \sigma_{3ij} \wedge \omega_{4kj}-\sum^{N}_{i=1}\int_{W_{7i}} \sigma_{3ii} \wedge \omega_{4ii}),
\end{equation}
where $Q_{1} = 2\kappa^{2} T_{5}$.

\section{The anomaly matching}
At the origin of the moduli space, 6d, $A_{N-1}$ (2, 0) field theory has the following form of anomaly when coupled to a background $SO(5)_{R}$ gauge field 1-form $A$, and in a general gravitational background \cite{C1a, C1b}.
\begin{equation}
I_{8}(N) = (N-1) I_{8}(1)+\frac{1}{24}(N^{3}-N)p_{2}(F). 	
\end{equation}
$I_{8}(1)$ is the anomaly polynomial for a single, free, (2, 0) tensor multiplet \cite{C2a, C2b}: 
\begin{equation}
I_{8}(1) = \frac{1}{48}\left[p_{2}(F)- p_{2}(R) +\frac{1}{4} (p_{1}(F)- p_{1}(R))^{2} \right].	
\end{equation}
$p_{2}(F)$ is the second Pontryagin class for the background $SO(5)_{R}$ field strength $F$:
\begin{equation}
	p_{2}(F) = \frac{1}{8}(\frac{i}{2\pi})^{4}\left[(trF^{2})\wedge(tr F^{2}) -2 trF^{4}   \right].
\end{equation}
At a generic point of the moduli space, the only massless degrees of freedom are $N-1$ copies of tensor multiplets giving rise to the anomaly of $(N-1) I_{8}(1)$. However, based on 't Hooft anomaly matching condition, the integration out of the massive degrees of freedom will produce the WZ term in the effective action, which will offer the missing $(N^{3}-N)p_{2}(F)/24$ part so that the total anomaly is still the same as before \cite{A}. In the following, we will show that the WZ term in (\ref{j}) could indeed give the $(N^{3}-N)p_{2}(F)/24$ part of the normal bundle anomaly.   

Turn on the background $SO(5)_{R}$ gauge field $A$ on $W_{6}$. $\partial W_{7} = W_{6}$, so $A$ could be smoothly extended to $W_{7}$. On $W_{7}$, we have gauge field $A^{ab}_{i} = -A^{ba}_{i}$, with $a, b = 1\cdots5$, $i = 1\cdots7$. In presence of the background field $A$, the pullback of the $S^{4}$ unite volume form on $W_{7}$ becomes 
\begin{eqnarray} 
\omega_{4}(\hat{\phi}, A) &=&  \frac{1}{2}e_{4}(\hat{\phi}, A) = \frac{1}{64\pi^{2}}\epsilon_{a_{1}\cdots a_{5}}[(D_{i_{1}}\hat{\phi})^{a_{1}} (D_{i_{2}}\hat{\phi})^{a_{2}}(D_{i_{3}}\hat{\phi})^{a_{3}} (D_{i_{4}}\hat{\phi})^{a_{4}} \nonumber \\ && -2F^{a_{1}a_{2}}_{i_{1}i_{2}}  (D_{i_{3}}\hat{\phi})^{a_{3}} (D_{i_{4}}\hat{\phi})^{a_{4}} + F^{a_{1}a_{2}}_{i_{1}i_{2}}F^{a_{3}a_{4}}_{i_{3}i_{4}} ] \hat{\phi}^{a_{5}} dx^{i_{1}}\wedge \cdots \wedge dx^{i_{4}},
\end{eqnarray}
$(D_{i}\hat{\phi})^{a} = \partial_{i} \hat{\phi}^{a}- A^{ab}_{i} \hat{\phi}^{b}$, $F^{ab}_{ij}$ is the field strength. $\hat{\phi}$ is a unite vector in the transverse space $R^{5}$. If $\omega_{4}(\hat{\phi}, A)$ represents the pullback of the 4-form field strength generated by the $i_{th}$ M5 brane on $W_{7j}$, $\hat{\phi}$ is determined by the relative position of $W_{6i}$ and $W_{7j}$. Especially, at the boundary of $W_{7j}$, $\hat{\phi}$ is simply determined by the relative position of $W_{6i}$ and $W_{6j}$ in the transverse space, i.e. 
\begin{equation}
\hat{\phi}^{a} = \frac{\phi^{a}_{i}- \phi^{a}_{j}}{|\phi_{i}- \phi_{j}|}, 	
\end{equation}
where $\phi^{a}_{i}$ is the vacuum expectation value of scalar field for the $i_{th}$ M5 brane. 

$e_{4}(\hat{\phi}, A)$ is the global angular form defined over the sphere bundle with fiber $S^{4}$ and base space $W_{7}$.
\begin{equation}
	d e_{4} = 0. 
\end{equation}
Under the $SO(5)$ transformation,
\begin{eqnarray} \label{k}
\hat{\phi}^{a} & \rightarrow &  \hat{\phi}^{a} + \Lambda^{ab} \hat{\phi}^{b} \nonumber \\ 
A^{ab}& \rightarrow & A^{ab} + d \Lambda^{ab} + [\Lambda, A]^{ab}.
\end{eqnarray}
$D \hat{\phi}^{a}$ and $F^{ab}$ transform covariantly under (\ref{k}), while $e_{4}(\hat{\phi}, A)$ is $SO(5)$ invariant. For the present problem, we have $N^{2}-N$ global angular forms $e_{4}(\hat{\phi}, A)$ with different $\hat{\phi}$ but the same $A$. Since $e_{4}$ is $SO(5)$ invariant, they can also be equivalently represented by $e_{4}(\hat{\phi}, A)$ with the same $\hat{\phi}$ but different $A$. $p_{2}(F)$ is the second Pontryagin class of a rank 5 real vector bundle, nevertheless, we still have
\begin{equation}
	p_{2}(F) = \chi (F)^{2}, 
\end{equation}
where $\chi (F)$ is the Euler class of a rank 4 subbundle with the orthogonal line bundle trivial. One can always choose particular $\hat{\phi}_{0}$ so that
\begin{equation}
	e_{4}(\hat{\phi}_{0}, A) = \chi (F)
\end{equation}
Actually, for such $\hat{\phi}_{0}$, $D \hat{\phi}_{0}= 0$, so $e_{4}(\hat{\phi}_{0}, A)$ reduces to the Euler class. We take this $\hat{\phi}_{0}$ as the standard and transform all of the angular forms into the form of $e_{4}(\hat{\phi}_{0}, \tilde{A})$, where $\tilde{A}$ are different connections defined on the same normal bundle. Just as the invariant polynomials, if $\tilde{A}$ and $\tilde{A}'$ are two different connections, 
\begin{equation}
	e_{4}(\hat{\phi}_{0}, \tilde{A})- e_{4}(\hat{\phi}_{0}, \tilde{A}') = de_{3}(\hat{\phi}_{0}, \tilde{A})- de_{3}(\hat{\phi}_{0}, \tilde{A}')= d R (\hat{\phi}_{0}, \tilde{A}, \tilde{A}'), 
\end{equation}
with $e_{3}$ the corresponding Chern-Simons forms \cite{C1a}.
\begin{equation}
R (\hat{\phi}_{0}, \tilde{A}, \tilde{A}') = - \frac{1}{32\pi^{2}}	\int^{1}_{0} dt \;\; \epsilon_{a_{1}\cdots a_{5}} [(D_{t}\hat{\phi}_{0})^{a_{1}}(D_{t}\hat{\phi}_{0})^{a_{2}}- F_{t}^{a_{1}a_{2}} ]\eta^{a_{3}a_{4}}\hat{\phi}_{0}^{a_{5}},
\end{equation}
where 
\begin{equation}
\eta = \tilde{A}- \tilde{A}', \;\;\;A_{t} = \tilde{A}' + t \eta,  \;\;\;F_{t} = dA_{t} - A_{t}^{2}, \;\;\; D_{t} = (d-A_{t}). 	
\end{equation}
$R (\hat{\phi}_{0}, \tilde{A}, \tilde{A}')$ is $SO(5)$ invariant. 
\begin{equation}
e_{3}(\hat{\phi}_{0}, \tilde{A})- e_{3}(\hat{\phi}_{0}, \tilde{A}')= R (\hat{\phi}_{0}, \tilde{A}, \tilde{A}'). 	
\end{equation}
For different connections, $e_{3}$ only differ by a $SO(5)$ invariant term.

Return to the original global angular form $e_{4}(\hat{\phi}, A)$, we will have 
\begin{equation}
e_{4}(\hat{\phi}, A) = 	\chi (F) + d \alpha (\hat{\phi}, A), 
\end{equation}
\begin{equation}
e_{4}(\hat{\phi}, A) \wedge	e_{4}(\hat{\phi}', A) = p_{2}(F) + d \beta (\hat{\phi}, \hat{\phi}', A), 
\end{equation}
where both $\alpha$ and $\beta$ are $SO(5)$ invariant.
\begin{equation}
p_{2}(F) =d [e_{3}(\hat{\phi}, A) \wedge	e_{4}(\hat{\phi}', A) - \beta (\hat{\phi}, \hat{\phi}', A)]. 
\end{equation}
By descent equations,
\begin{equation}
	\delta [e_{3}(\hat{\phi}, A) \wedge	e_{4}(\hat{\phi}', A)] = \delta p_{2}^{0}(A) = d p_{2}^{1}(A). 
\end{equation} 
Now, consider the $SO(5)$ gauge transformation of (\ref{j}). For each term,
\begin{eqnarray}
\delta[	\frac{1}{6} \int_{W_{7}} \sigma_{3}(\hat{\phi}, A) \wedge \omega_{4}(\hat{\phi}', A) ] &=& \delta[ \frac{1}{24} \int_{W_{7}} e_{3}(\hat{\phi}, A) \wedge	e_{4}(\hat{\phi}', A)] \nonumber \\ &=&\frac{1}{24} \int_{W_{7}}d p_{2}^{1}(A) = \frac{1}{24} \int_{W_{6}} p_{2}^{1}(A). 
\end{eqnarray}
There are totally $N^{3}- N$ such terms, so $\Gamma_{WZ}$ could indeed reproduce the $(N^{3}- N)p_{2}(F)/24$ part of the anomaly.

If the $SU(N)$ group is broken to some subgroup like $U(N_{1})\times U(N_{2})\times SU(N_{3})$ with $N_{1}+N_{2}+N_{3} = N$, the deficit of the anomaly produced by massless degrees of freedoms is $(N^{3}- N_{1}^{3}- N_{2}^{3}- N_{3}^{3})p_{2}(F)/24$. Corresponding, $\Gamma_{WZ}$ will contain $N^{3}- N_{1}^{3}- N_{2}^{3}- N_{3}^{3}$ terms exactly compensating the deficit.

\section{The degrees of freedom in M5 branes producing the WZ term}

The supergravity interaction between D-branes is pairwise. This is consistent with the fact that the WZ term for N D-branes in a generic Coulomb branch could be written as the sum of $N(N-1)/2$ terms labeled by $(ij)$ index \cite{D, E}. On the other hand, the $\int A_{3}\wedge F_{4}$ term in the action of M5 branes gives a triple interaction. That is, three M5 branes could interact simultaneously. The supergravity interaction for D-branes is produced by open strings connecting two D-branes. Similarly, one may expect that the triple interaction $\int A_{3}\wedge F_{4}$ could be produced by open M2 branes connecting three M5 branes. 

Another example of $N^{3}$ interaction is given by M theory compactified on a Calabi-Yau threefold with M5-branes wrapping 4-cycles, giving rise to $N = 1$ 5d supergravity along with the chiral strings \cite{G, H}. In the bulk, we have Chern-Simons term $C_{1} \wedge dC_{1} \wedge dC_{1}$, while in the worldsheet of chiral strings, $\int C_{1} \wedge dC_{1}$ may exist \cite{A}. These $N^{3}$ degrees of freedom in entropy are explained as states living at the triple-intersection of M5 branes \cite{III, I}.  

In \cite{K} and more recently, \cite{L}, the $1/4$ BPS objects in the Coulomb phase of the ADE-type 6d $(2, 0) $ superconformal theories are explored. They are made of waves on selfdual strings and junctions of selfdual strings. Especially, in \cite{L}, it is shown that the number of $1/4$ BPS objects matches exactly one third of the anomaly constant $c_{G} = d_{G}h_{G}$ for all ADE types, which strongly indicates that the anomaly may be produced by these $1/4$ BPS objects. In $A_{N-1}$ case, there are $N(N-1)/2$ $1/2$ BPS selfdual strings with tension $T_{ij}\propto |\phi_{i}- \phi_{j}|$. On each selfdual string, there are left and right $1/4$ BPS waves. Turning on these BPS waves, we get $N(N-1)$ $1/4$ BPS objects. For every three M5 branes $ijk$, $1/4$ BPS junction exists. The tension of the string junctions is characterized by $(|\phi_{i}- \phi_{j}|, |\phi_{j}- \phi_{k}|, |\phi_{k}- \phi_{i}|)$. The junction forms a dual lattice to the triangle $\Delta_{ijk}$, if one indentify the $SO(5)$ in $W_{6}$ with the $SO(5)$ in the transverse space. For such configuration, the tension of selfdual strings is balanced and the junction is $1/4$ BPS. There are totally $N(N-1)(N-2)/3$ such objects because of the junction and anti-junction. Altogether, the $1/4$ BPS objects on $N$ M5 branes in a generic Coulomb branch is $N(N^{2}-1)/3$.

Let us rewrite (\ref{j}) in a more symmetric way.
\begin{equation}\label{n}
2\kappa^{2} \Gamma_{WZ} = \sum_{i\neq j,j\neq k, k\neq i} \Omega_{ijk} + \sum_{i\neq j}\Omega_{ij}, 
\end{equation}
where 
\begin{equation}\label{l}
\Omega_{ijk} = \frac{Q_{1}^{3}}{6}[\int_{W_{7i}} (\sigma_{3ji} \wedge \omega_{4ki}+ \sigma_{3ki} \wedge \omega_{4ji})+ \int_{W_{7j}} (\sigma_{3ij} \wedge \omega_{4kj}+ \sigma_{3kj} \wedge \omega_{4ij}) +  \int_{W_{7k}} (\sigma_{3ik} \wedge \omega_{4jk}+ \sigma_{3jk} \wedge \omega_{4ik})   ]	
\end{equation}
\begin{equation}\label{m}
\Omega_{ij} = \frac{Q_{1}^{3}}{6}[\int_{W_{7i}} (\sigma_{3ji} \wedge \omega_{4ii}+ \sigma_{3ii} \wedge \omega_{4ji}+ \sigma_{3ji} \wedge \omega_{4ji})+ \int_{W_{7j}} (\sigma_{3ij} \wedge \omega_{4jj}+ \sigma_{3jj} \wedge \omega_{4ij}+ \sigma_{3ij} \wedge \omega_{4ij})   ]. 		
\end{equation}
It seems that junction and anti-junction may produce the term $\Omega_{ijk}$, while left and right waves on selfdual strings could give $\Omega_{ij}$. Recall that in D-brane case, the WZ term arising from the integration out of massive fermions $\psi_{ij}$ is expressed in terms of the vector $\phi_{i}- \phi_{j}$ \cite{D, E}; here, the WZ term produced by string junctions ($ijk$) could be calculated from the vectors $(\phi_{i}- \phi_{j}, \phi_{j}- \phi_{k}, \phi_{k}- \phi_{i})$. When $i = k$, the three string junctions degenerate to one selfdual string with tension $T_{ij}\propto |\phi_{i}- \phi_{j}|$ and the other tensionless selfdual string perpendicularly ending on it. So, in some sense, selfdual string with waves is a degeneration of the string junction. $\Omega_{ij} = \frac{1}{2}(\Omega_{iji}+ \Omega_{jij})$.

Except for $\int_{W_{7}} A_{3}\wedge F_{4}$, M5 brane action contains another term $\int_{W_{6}} db_{2} \wedge A_{3} =-\int_{W_{6}} b_{2} \wedge F_{4} = \int_{W_{6}} H_{3} \wedge A_{3}$. Now suppose the vacuum expectation value of $b_{2}$ on the $i_{th}$ M5 brane is $b_{2i}$, $\int_{W_{6}} H_{3} \wedge A_{3}$ part of the WZ term should also enter into the low energy effective action, although it does not contribute to the anomaly since $d H_{3} = 0$. In \cite{J}, based on the supergravity calculation, it is shown that 
\begin{equation}\label{asdfg}
\Gamma_{H} = \int_{W_{6}} H_{3} \wedge A_{3} \propto  - \sum_{_{i\neq j}} \int_{W_{6}} b_{2ij} \wedge \omega_{4ij} = \sum_{_{i\neq j}} \int_{W_{6}} H_{3ij} \wedge \sigma_{3ij}, 	
\end{equation}
where $b_{2ij} = b_{2i} - b_{2j}$, $H_{3ij} = d b_{2ij} = H_{3i}-H_{3j}$. This is the typical pairwise interaction. The reduction of $\int_{W_{6}} H_{3} \wedge A_{3}$ on $S^{1}$ gives $\int_{W_{5}} F_{2} \wedge A_{3}$, the WZ term of the 5d SYM theory. In 5d SYM theory, $\int_{W_{5}} F_{2} \wedge A_{3}$ is generated by the integration out of massive fermions coming from the selfdual strings wrapping $S^{1}$, so it is quite possible that (\ref{asdfg}) is produced by $1/2$ BPS selfdual strings.

The non-abelian part of the R-symmetry anomaly are all accounted for by $1/4$ BPS objects. R-symmetry anomaly and Weyl anomaly are related by supersymmetry. In \cite{M}, the conformal anomaly of 6d (2, 0) SCFT of $A_{N-1}$ type is calculated as
\begin{equation}
	A_{2, 0}(N) = (N-1) A_{tens} + (N^{3}-N) A, 
\end{equation}
where $A_{tens}$ is the conformal anomaly of the free $(2,0)$ tensor multiplet. It is expected that the $1/4$ BPS objects could produce $(N^{3}-N) A$, if they could give the corresponding part in R-symmetry anomaly. Note that $N-1$ 1/2 BPS massless particles and $N(N-1)(N-2)/3$ junctions of selfdual strings always contribute to the anomaly and entropy. However, the $N(N-1)/2$ selfdual strings have no contribution to the anomaly nor entropy unless the BPS waves are turned on thus the supersymmetry is reduced to $1/4$. Once the selfdual strings become $1/4$ BPS, the anomaly polynomial of them is the same as that of the string junctions, since the $1/4$ BPS selfdual strings could be taken as the degeneration of the string junctions. Then the question is why the $1/2$ BPS selfdual strings have no contribution to the anomaly nor the entropy. In $N=4$ SYM theory, $1/4$ BPS states arising from string junctions ending on three D3 branes also exist \cite{N}, however, the anomaly and entropy are both give by $1/2$ BPS particles. In some sense, the $1/4$ BPS states could be taken as the bound states of the $1/2$ BPS states, so it is likely that for $N=4$ SYM theory, $1/2$ BPS states are fundamental, while for $6d$ $(2, 0)$ theory, it is the bound states which are dominating.

The general form of the anomaly for a 6d $(2, 0)$ SCFT of the ADE type G is  
\begin{equation} 
A_{2, 0} = r_{G} A_{tens} + c_{G} A_{X}, 	
\end{equation}
where $c_{G} = d_{G}h_{G} = r_{G} h_{G} (h_{G}+1)$. $r_{G}$, $d_{G}$ and $h_{G}$ are the rank, the dimension, and the Coxeter number of the Lie algebra of type G. The theory contains $r_{G}$ $1/2$ BPS massless particles, $r_{G} h_{G}$ $1/2$ BPS selfdual strings, and $c_{G}/3$ $1/4$ BPS objects. The anomaly of the single M5 brane does not have the $A_{X}$ part, so $r_{G}$ $1/2$ BPS massless particles only contribute to $A_{tens}$. Then $A_{X}$ should be generated by $1/2$ BPS selfdual strings or $1/4$ BPS objects. If one wants to interpret it in terms of selfdual strings, each selfdual string should give the anomaly of $(h_{G}+1) A_{X}$, which in $SU(N)$ case, is $(N+1) A_{X}$. It is difficult to explain this $h_{G}+1$ factor. Otherwise, since the total number of the $1/2$ BPS states is $d_{G}$, they can account for the $A_{X}$ part if each one contributes $h_{G} A_{X}$. This looks more reasonable, but the problem is that the $r_{G}$ $1/2$ BPS massless particles will contribute to $A_{tens}$ as well as $A_{X}$. The most natural possibility is that $A_{X}$ is produced by $1/4$ BPS objects, which are intrinsically three selfdual string junctions.

Finally, notice that for $N=4$ SYM theory, the anomaly takes the form $A_{4} = (N^{2}-1)A_{vec}$, where $A_{vec}$ is the anomaly of a free vector multiplet. The anomaly is not renormalized from weak to strong coupling, so we can calculate it from the free field value. Besides, $N^{2}-1$ elements in the Lie algebra give the same contribution to the anomaly, indicating that they are allowed to transform into each other. On the other hand, for $6d$ (2, 0) SCFT, the anomaly polynomial is of the form $A_{2, 0} = (N-1) A_{tens} + (N^{3}-N) A_{X}$ other than $(N^{3}-1) A$, which seems indicate that there are something special about the non-abelian part.

$6d$ $(2, 0)$ SCFT compactified on $S^{1}$ gives $5d$ SYM theory. Selfdual strings wrapping on $S^{1}$ become $1/2$ BPS particles. The unwrapped selfdual strings and $1/4$ BPS string junctions in $6d$ descend to the corresponding string-like objects in $5d$ \cite{K, L}. String junctions may also appear as point-like particles in the compactified theory. Consider the $6d$ SCFT compactified on a Riemann surface $\Sigma_{g}$ with $g>1$ \cite{P, Q, R}, the T part of $\Sigma_{g}$ is the natural place for string junctions to wrap. $\Sigma_{g}$ is built from $2(g-1)$ $T_{N}$ blocks and $3(g-1)$ $I_{N}$ blocks. $T_{N}$ and $I_{N}$ are spheres with $3$ and $2$ full punctures respectively. The dimension of the Coulomb branch for $T_{N}$ and $I_{N}$ are  
\begin{equation}
	d_{c}T_{N} = \frac{(N-1)(N-2)}{2}, \;\;\;\;\;\;d_{c}I_{N} = N-1. 
\end{equation}
The effective number of vector multiplets for $T_{N}$ and $I_{N}$ are    
\begin{equation}
	n_{v}T_{N} = \frac{2N^{3}}{3}-\frac{3N^{2}}{2}-\frac{N}{6}+1, \;\;\;\;\;\; n_{v}I_{N} = N^{2}-1. 
\end{equation}
Note that 
\begin{equation}
	2(g-1) d_{c}T_{N} + 3(g-1) d_{c}I_{N} = (g-1)(N^{2}-1)
\end{equation}
and 
\begin{equation}
	2(g-1) n_{v}T_{N} + 3(g-1) n_{v}I_{N} = (g-1)(\frac{4N^{3}}{3}-\frac{N}{3}-1)
\end{equation}
are the dimension of the Coulomb branch and the effective number of vector multiplets for the $\Sigma_{g}$ theory. Especially, when $g=1$, $\Sigma_{1}$ is simply constructed from one $I_{N}$. The degrees of freedom arising from strings could be calculated as
\begin{equation}
	n_{T_{N} } =n_{v}T_{N} -d_{c}T_{N}  = 4 C^{3}_{N}, \;\;\;\;\;\; n_{I_{N} }= n_{v}I_{N} - d_{c}I_{N}= 2C^{2}_{N}. 
\end{equation}
$n_{T_{N} }$ could be naturally accounted for by string junctions\footnote{String junctions could give $2 C^{3}_{N}$, but the extra factor $2$ is a little difficult to explain.}, while $n_{I_{N} }$ is associated with the selfdual strings. In the generic point of the Coulomb branch of $T_{N}$, the Seiberg-Witten curve is a Riemann surface with $(N-1)(N-2)/2$ genus and $3N$ simple punctures \cite{Q, SSB}. In that case, $n_{T_{N} }$ could also be explained as the number of M2 branes with two boundaries. However, at the origin of the moduli space, the only nontrivial configurations are M2 branes with three boundaries. The anomaly polynomial $I_{6}(N)$ in $4d$ is obtained from $I_{8}(N)$ in $6d$ by the integration over $\Sigma_{g}$ \cite{S, T}. Both of them have a $N^{3}$ scaling part.

\section {WZ term from the integration of massive fermions}

The WZ term could be derived by a 1-loop calculation in field theory. For $SU(N)$ gauge theories, this has been done in \cite{D, E, F}. In Coulomb branch, fermions get mass due to the Yukawa coupling. The integration of the fermion loop gives WZ terms in the low energy effective action. We don't know the structure of the $6d$ $(2, 0)$ field theory. A recent calculation on scattering amplitudes \cite{U} indicates that an interacting $6d$ Lagrangian with classical $OSp(8|4)$ symmetry cannot be constructed using only $(2, 0)$ tensor multiplets, even if the Lagrangian is non-local. A Lagrangian description may exist, however, if one includes additional degrees of freedom, for example, the selfdual strings. It is possible that the $6d$ $(2, 0)$ theory may be quite different from the ordinary field theories, and so, the WZ term in the effective action should also be calculated in a peculiar way. Alternatively, if the theory is constructed as a normal QFT containing fermionic degrees of freedom and the corresponding Dirac operators, then at a generic Coulomb branch, the WZ term may be calculated by a standard fermion-loop integration. In both cases, the Hopf-Wess-Zumino term should be generated to compensate the anomaly deficit. In the following, we will discuss the WZ term for the second possibility, especially, for the 3-algebra valued tensor multiplet.

For the calculation of the WZ term, the input from the field theory is the Dirac operator involving Yukawa couplings as well as the gauge couplings. For $SU(N)$ gauge theories, the Yukawa coupling is $\Gamma_{I}[X^{I}, \psi]$. In Coulomb branch, it becomes $\Gamma_{I}(\phi^{I}_{i}-\phi^{I}_{j})\psi_{ij}$, giving mass $|\phi_{i}-\phi_{j}|$ to $\psi_{ij}$. $|\phi_{i}-\phi_{j}|$ is the length of the string connecting the $i_{th}$ and the $j_{th}$ D-branes. On the other hand, the Lagrangian of the M2 branes has a sextic potential \cite{X1,X2,X3,Y}, so the fermion mass scales as the area other than the length, reflecting the fact that M2 branes are connected by M2 branes other than strings \cite{V, W}. M5 branes are also connected by M2 branes, so we may expect that the fermion mass will give the area of the M2 branes connecting different M5 branes. However, there is a difference. The M2 branes connecting parallel M2 branes are totally located in the transverse space. The endpoint is simply a point. As a result, the Yukawa coupling in BLG model takes the form of $\Gamma_{I}\Gamma_{J}[X^{I},X^{J}, \psi]$ \cite{X1,X2,X3}, where $X^{I}$ and $X^{J}$ carry the transverse index. Conversely, the M2 branes connecting parallel M5 branes have one dimension living in the worldvolume of M5. The endpoint is a string. Correspondingly, we may have $\Gamma_{\mu}\Gamma_{I}[C^{\mu},X^{I}, \psi]$, where $\mu = 0 \cdots 5$, $I = 6 \cdots 10$. 

Actually, in \cite{Z}, an attempt to find the $6d$ $(2, 0)$ theory with the 3-algebra structure has already been made. It was shown that for the closure of the supersymmetry, an additional vector $C^{\mu}$ must be introduced, while the 3-brackets appearing in the equations of motion always take the form $[C^{\mu},A, B]$. Later, in \cite{a, b}, the equations of motion found in \cite{Z} get a natural interpretation as the supersymmetric gauge field equations in loop space. $C^{\mu}$ is associated with the vector tangential to the loop in the worldvolume of M5 branes.

The field content in \cite{Z} includes the tensor multiplet composed by $X^{I}$ with $I = 6 \cdots 10$, $\psi$, and $H^{\mu\nu\lambda}$ with $\mu, \nu, \lambda = 0 \cdots 5$, an auxiliary gauge field $A^{\mu}$, and a vector field $C^{\mu}$. $X^{I}$, $\psi$, $H^{\mu\nu\lambda}$, and $C^{\mu}$ take values in a vector space $\Lambda$ with the basis $t^{a}$, i.e. $X^{I} = X^{I}_{a}t^{a}$, etc. As a 3-algebra, $\Lambda$ has an associated Lie algebra $\textit{g}_{\Lambda}$ spanned by the transformations $[t^{a},t^{b}, *]$, where $*$ stands for an arbitrary element of $\Lambda$. $A^{\mu}$ takes values in $\textit{g}_{\Lambda}$. $A^{\mu}X^{I} = A^{\mu}_{ab}[t^{a},t^{b},X^{I}]$, etc. $A^{\mu}$ and $H^{\mu\nu\lambda}$ are related by $F^{\mu\nu}_{ab}[t^{a},t^{b}, *] = [C_{\lambda}, H^{\mu\nu\lambda}, *]$, with $F^{\mu\nu}_{ab}$ the field strength of $A^{\mu}_{ab}$. So, for the given $C_{\lambda}$, $F^{\mu\nu}$ is actually a transgression of $H^{\mu\nu\lambda}$ \cite{a}.

The equations of motion for 3-algebra valued (2, 0) tensor multiplets found in \cite{Z} are
\begin{eqnarray}\label{18} 
&&[C^{\mu},C^{\nu}, *]=0,
 \\
&& \nabla_{\nu}C^{\mu} = 0,   \\
&& [C^{\rho},\nabla_{\rho} X^{I}, *] = 0,\;\;\ [C^{\rho},\nabla_{\rho} \psi, *] = 0,\;\;\ [C^{\rho},\nabla_{\rho} H_{\mu\nu\lambda}, *] = 0, \label{26}\\
&&  \tilde{F}^{\mu\nu} - [C_{\lambda}, H^{\mu\nu\lambda}, *] = 0, \label{19} \\
&& \Gamma_{\mu} \nabla^{\mu} \psi +  [C^{\mu}, X^{I}, \Gamma_{\mu} \Gamma_{I} \psi] = 0, \label{20}	\\
&& \nabla^{2}X^{I} -\frac{i}{2}[C^{\mu}, \bar{\psi}, \Gamma_{\mu} \Gamma^{I} \psi]+ [C^{\mu}, X^{J},  [C_{\mu},X_{J},  X^{I} ]] = 0,\\
&& \nabla_{[\mu}H_{\nu\kappa\lambda]} +  \frac{1}{4} \epsilon_{\mu\nu\kappa\lambda\sigma\tau}[C^{\sigma}, X^{I},\nabla^{\tau}X_{I}]+ \frac{i}{8}\epsilon_{\mu\nu\kappa\lambda\sigma\tau}[C^{\sigma}, \bar{\psi},  \Gamma^{\tau} \psi]=0, \label{21}
\end{eqnarray}
where $\nabla^{\mu}=\partial^{\mu}-i A^{\mu}$, $X^{I}, \psi, H_{\mu\nu\lambda} \in \Lambda$, $C^{\mu}\in \bar{\Lambda}$. $\forall \; C_{1}, C_{2} \in \bar{\Lambda}$, $[C_{1},C_{2},*] = 0$. 
\begin{equation}
	\tilde{F}^{\mu\nu} = \partial^{\nu}A^{\mu} - \partial^{\mu}A^{\nu} +[A^{\nu}, A^{\mu}]. 
\end{equation}
The supersymmetry transformations are \cite{Z}
\begin{eqnarray}\label{27} 
&& \delta X^{I}	  = i \bar{\varepsilon} \Gamma^{I} \psi, 
 \nonumber \\
&& \delta \psi = \Gamma_{\mu}\Gamma_{I} \nabla^{\mu} X^{I} \varepsilon +\frac{1}{12} \Gamma_{\mu\nu\lambda}H^{\mu\nu\lambda} \varepsilon -\frac{1}{2} \Gamma_{IJ}\Gamma_{\lambda} [X^{I},X^{J},C^{\lambda}]\varepsilon, \nonumber
  \\
&& \delta H_{\mu\nu\lambda} = 3i \bar{\varepsilon} \Gamma_{[\mu\nu}\nabla_{\lambda]}\psi + i \bar{\varepsilon} \Gamma_{I} \Gamma_{\mu\nu\lambda\kappa}[X^{I},\psi,C^{\kappa}], \nonumber
\\
&& \delta A_{\mu} = i \bar{\varepsilon} \Gamma_{\mu\lambda} [C^{\lambda},\psi,*], \nonumber
\\
&& \delta C_{\mu} = 0. 
\end{eqnarray}

When $C^{\mu}=0$, (\ref{18})-(\ref{21}) reduce to the equations of motion for free tensor multiplets. We may take $C^{\mu}$ as a vector associated with the selfdual strings. The length of $C^{\mu}$ may characterize the length of strings. When the strings shrink to points, which are described by tensor multiplets, the interaction disappears. A particular $C^{\mu}$ corresponds to a particular set of selfdual strings with the given length and the orientation. We want to take $C^{\mu}$ as the new degrees of freedom added, so we will not specify it. The path integral may cover all possible configurations of $C^{\mu}$.

Now, consider the Coulomb branch of the theory. The supersymmetry transformation (\ref{27}) suggests that the vacuum configuration is given by constant $X^{I}$ satisfying $[C^{\mu}, X^{I}, X^{J}] = 0$. Choose a maximal subspace $\Lambda_{0}$, $\Lambda_{0} \subset \Lambda$, $\forall A, B \in \Lambda_{0}$, $\forall C \in \bar{\Lambda}$, $[C, A, B]=0$. $\Lambda = \Lambda_{0}\oplus \Lambda_{1}$, there is a special set of basis $\left\{t^{1}\cdots t^{M} \right\}$ for $\Lambda_{1}$, $\forall C \in \bar{\Lambda}$, $\forall A \in \Lambda_{0}$, $[C, A, t^{m}] \propto t^{m}$. $\Lambda_{0}$ and $\left\{t^{1}\cdots t^{M} \right\}$ could be taken as the Cartan subalgebra and the roots respectively. Suppose the vacuum expectation value of $X^{I}$ is given by $\bar{X}^{I}$, $\bar{X}^{I} \in \Lambda_{0}$. $[C_{\mu}, \bar{X}^{I}, t^{m}] = \phi^{I}_{m \mu} t^{m}$. Similarly, suppose the vacuum expectation value of $H^{\mu\nu\lambda}$ is $\bar{H}^{\mu\nu\lambda}$, $\bar{H}^{\mu\nu\lambda} \in \Lambda_{0}$, then $[C_{\lambda}, \bar{H}^{\mu\nu\lambda}, t^{m}] = f^{\mu\nu}_{m} t^{m}$.

Plug the vacuum expectation values into the Dirac equation for fermions. Let $\psi^{m}$ denote fermions taking values in the root $t^{m}$. From (\ref{20}),   
\begin{equation}\label{re}
	\Gamma_{\mu} \nabla^{\mu} \psi^{m} + \phi^{I}_{m \mu} \Gamma^{\mu} \Gamma_{I} \psi^{m}  = 0. 
\end{equation}
Besides,
\begin{equation}
[C_{\lambda},  \bar{H}^{\mu\nu\lambda},\Gamma_{\mu}\Gamma_{\nu} \psi^{m}] = f^{\mu\nu}_{m} \Gamma_{\mu}\Gamma_{\nu} \psi^{m}.
\end{equation}
The form of the Dirac operator in (\ref{re}) is quite natural, so it may also appear in other models for $6d$ $(2, 0)$ theories, for example, \cite{cccc}. $\psi^{m}$ are the $6d$ anti-chiral fermions. Written as the $11d$ Majorana spinors, $\Gamma_{7} \psi^{m} = - \psi^{m}$, where $\Gamma_{7} = \Gamma_{012345}$. Just as that in \cite{D, E}, the WZ term could be written as
\begin{equation}
\Gamma_{m}=  \texttt{Tr}  \left\{\texttt{ln} \: [i \Gamma_{0}\Gamma_{\mu}\partial^{\mu} +  \Gamma_{0} \Gamma_{\mu}A_{m}^{\mu} +  i \Gamma_{0} \Gamma^{\mu} \Gamma_{I} \phi^{I}_{m \mu} ]\frac{1-\Gamma_{7}}{2} \right\},	
\end{equation}
\begin{equation}\label{4}
\frac{\delta \Gamma_{m}}{\delta  \phi^{I}_{m \mu}(x)} = Sp  [\left\langle x \right| \frac{ 1}{i \Gamma_{\nu}\partial^{\nu} + \Gamma_{\nu}A_{m}^{\nu} +  i \Gamma^{\nu} \Gamma_{I}  \phi^{I}_{m \nu}}  \left|  x\right\rangle i \Gamma^{\mu} \Gamma_{I} (\frac{1-\Gamma_{7}}{2})],	
\end{equation}
where $Sp$ is the trace in spinor indices. WZ term comes from the imaginary part of the effective action. Taking the difference of (\ref{4}) with its complex conjugate,
\begin{eqnarray}\label{5} 
\frac{\delta Im \Gamma_{m}}{\delta  \phi^{I}_{m \mu}(x)}	&=&  - \frac{1}{2} Sp  [\left\langle x \right| \frac{ 1}{i \Gamma_{\nu}\partial^{\nu} + \Gamma_{\nu}A_{m}^{\nu} + i \Gamma^{\nu} \Gamma_{I}  \phi^{I}_{m \nu}}  \left|  x\right\rangle  \Gamma^{\mu} \Gamma_{I} \Gamma_{7}]
\nonumber \\
&=& -\frac{1}{2} Sp  [\left\langle x \right| \frac{ 1}{\slash{D}}  \left|  x\right\rangle  \Gamma^{\mu} \Gamma_{I} \Gamma_{7}]\nonumber \\
&=& - \frac{1}{2} Sp  [\left\langle x \right| \frac{ \slash{D}}{\slash{D}^{2}}  \left|  x\right\rangle  \Gamma^{\mu} \Gamma_{I} \Gamma_{7}].
\end{eqnarray} 
(\ref{5}) could be expanded as the sum of the terms proportional to $1$ or $Tr(\Gamma_{M_{1}}\cdots \Gamma_{M_{k}})$, where $M_{1},\ldots, M_{k}$ are distinct indices. Note that $Tr(\Gamma_{M_{1}}\cdots \Gamma_{M_{k}})=0$ unless $k=11$, so we need to extract the term proportional to $Tr(\Gamma_{0}\cdots \Gamma_{10}) \propto \epsilon_{0\cdots10}$. In the numerator, $i \Gamma^{\nu} \Gamma_{I}  \phi^{I}_{m \nu}$ in $\slash{D}$ will be kept, while in the denominator,
\begin{equation}
\slash{D}^{2}= - \partial^{2} + \phi^{I}_{m \mu} \phi^{\mu}_{m I} + \frac{i}{2}\Gamma_{\mu}\Gamma_{\nu}f_{m}^{\mu\nu}- \Gamma_{\mu}\Gamma^{\nu}\Gamma_{I}\partial^{\mu}\phi^{I}_{m \nu}+ \Gamma^{\mu\nu}\Gamma_{IJ} \phi^{I}_{m\mu}\phi^{J}_{m\nu}+\cdots,
\end{equation}
where $f_{m}^{\mu\nu}=\partial^{\mu}A_{m}^{\nu}- \partial^{\nu}A_{m}^{\mu}+[A_{m}^{\mu}, A_{m}^{\nu}]$, 
\begin{equation}
\frac{1}{\slash{D}^{2}}=-\sum^{\infty}_{n=0}\frac{[\frac{i}{2}\Gamma_{\mu}\Gamma_{\nu}f_{m}^{\mu\nu}-\Gamma_{\mu}\Gamma^{\nu}\Gamma_{I}\partial^{\mu}\phi^{I}_{m \nu}+ \Gamma^{\mu\nu}\Gamma_{IJ} \phi^{I}_{m\mu}\phi^{J}_{m\nu}+ \cdots]^{n}}{(\partial^{2}- \phi^{I}_{m \mu} \phi^{\mu}_{m I})^{n+1}}.
\end{equation}
The integral that needs to be performed is
\begin{eqnarray}
\left\langle x \right| \frac{ 1}{(\partial^{2}- \phi^{I}_{m \mu} \phi^{\mu}_{m I})^{n+1}}  \left|  x\right\rangle 	&=& (-1)^{n+1} \int \frac{d^{d}p}{(2 \pi)^{d}} \frac{1}{(p^{2}+ \phi^{I}_{m \mu} \phi^{\mu}_{m I})^{n+1}}
\nonumber \\
&=& \frac{i \Gamma(n+1-\frac{d}{2})}{(2\sqrt{\pi})^{d}\Gamma(n+1)} \frac{1}{(\sqrt{\phi^{I}_{m \mu} \phi^{\mu}_{m I}})^{2n+2-d}}.
\end{eqnarray} 
The constraint $[C^{\mu}, \nabla_{\mu}\psi^{m}, *] = 0$ reduces the dynamics from $6d$ to $5d$. If $C^{\mu}$ is taken as the vector tangential to the selfdual string, the constraint means that the physical momentum of the string is along the transverse direction. As a result, $d=5$.

For $n=4$, $2n+2-d = 5$. 
\begin{equation}
\frac{\delta Im \Gamma_{m}}{\delta  \phi^{I}_{m \mu}(x)} \propto \frac{ Sp \left\{ [\frac{i}{2}\Gamma_{\rho}\Gamma_{\nu}f_{m}^{\rho\nu}-\Gamma_{\rho}\Gamma^{\nu}\Gamma_{J}\partial^{\rho}\phi^{J}_{m \nu}+ \Gamma^{\rho\nu}\Gamma_{KJ} \phi^{K}_{m \rho}\phi^{J}_{m\nu}+ \cdots]^{4} \phi^{L}_{m \lambda} \Gamma^{\lambda} \Gamma_{L}  \Gamma^{\mu} \Gamma_{I} \Gamma_{7}\right\}}{(\sqrt{\phi^{I}_{m \mu} \phi^{\mu}_{m I}})^{5}}
\end{equation}
The term containing one $f_{m}^{\rho\nu}$ and three $\partial^{\rho}\phi^{J}_{m \nu}$ is
\begin{eqnarray}\label{31}
\frac{\delta L_{1}}{\delta  \phi^{I}_{m \mu}(x)} 	&\propto & - \frac{ f_{m\rho\nu}  \partial_{\rho_{1}}\phi^{J_{1}}_{m \nu_{1}}\partial_{\rho_{2}}\phi^{J_{2}}_{m \nu_{2}}\partial_{\rho_{3}}\phi^{J_{3}}_{m \nu_{3}}\phi^{L}_{m \lambda}}{(\sqrt{\phi^{I}_{m \mu} \phi^{\mu}_{m I}})^{5}}  \nonumber \\
&& Sp[\Gamma^{\rho}\Gamma^{\nu} \Gamma^{\rho_{1}}\Gamma^{\nu_{1}}\Gamma^{\rho_{2}}\Gamma^{\nu_{2}}\Gamma^{\rho_{3}}\Gamma^{\nu_{3}} \Gamma^{\lambda}  \Gamma^{\mu}  \Gamma_{7}\Gamma_{J_{1}} \Gamma_{J_{2}}\Gamma_{J_{3}}\Gamma_{L}\Gamma_{I} ]  
\end{eqnarray} 
The counterpart of $L_{1}$ in the low energy effective action is $L'_{1} \propto \int_{W_{6}} H_{3} \wedge A_{3}$,
\begin{equation}\label{33}
	\frac{\delta L'_{1}}{\delta  \phi^{I}_{m }(x)}  \propto \frac{\epsilon^{\rho\nu\mu\rho_{1}\rho_{2}\rho_{3}}\epsilon_{J_{1}J_{2}J_{3}LI}h_{m \rho\nu\mu}
\partial_{\rho_{1}}\phi^{J_{1}}_{m }\partial_{\rho_{2}}\phi^{J_{2}}_{m }\partial_{\rho_{3}}\phi^{J_{3}}_{m}	\phi^{L}_{m}}{(\sqrt{\phi^{I}_{m } \phi_{m I}})^{5}}, 
\end{equation}
Here $m$ denotes a particular root. If $m \sim (i, j)$, $\phi^{I}_{m } = \phi^{I}_{i } - \phi^{I}_{j } = \phi^{I}_{ij }$, $h_{m \rho\nu\mu} = h_{i \rho\nu\mu}- h_{j \rho\nu\mu} = h_{ij \rho\nu\mu}$. In general, one may expect that $\phi^{I}_{\mu}$ could be expanded as $\phi^{I}_{\mu} = \sum_{k} c_{k\mu} a_{k}^{I}$, where $c_{k\mu}$ and $a_{k}^{I}$ are vectors along the longitudinal and transverse directions respectively. The simplest situation is $\phi^{J}_{m \nu} = c_{m \nu} \phi_{m}^{J}$. $\partial_{\rho}\phi^{J}_{m \nu} = \partial_{\rho}c_{m \nu} \phi_{m}^{J} + c_{m \nu}\partial_{\rho}\phi^{J}_{m } =  c_{m \nu}\partial_{\rho}\phi^{J}_{m }$. Therefore, $\partial_{\rho_{3}}\phi^{J_{3}}_{m \nu_{3}}\Gamma^{\rho_{3}}\Gamma^{\nu_{3}} \Gamma^{\lambda} \phi^{L}_{m \lambda} =  |c_{m}|^{2} \Gamma^{\rho_{3}}\partial_{\rho_{3}}\phi^{J_{3}}_{m}\phi^{L}_{m}$. Similarly, $\partial_{\rho_{1}}\phi^{J_{1}}_{m \nu_{1}}\partial_{\rho_{2}}\phi^{J_{2}}_{m \nu_{2}}\Gamma^{\rho_{1}}\Gamma^{\nu_{1}}\Gamma^{\rho_{2}}\Gamma^{\nu_{2}} \sim  -|c_{m}|^{2}\partial_{\rho_{1}}\phi^{J_{1}}_{m }\partial_{\rho_{2}}\phi^{J_{2}}_{m }\Gamma^{\rho_{1}}\Gamma^{\rho_{2}}$, where we have neglected the term involving $c^{ \nu}_{m}\partial_{\nu}\phi^{J}_{m } $. (\ref{31}) is simplified to 
\begin{equation}\label{32}
\frac{\delta L_{1}}{\delta  \phi^{I}_{m }(x)} \propto  \frac{  f_{m\rho\nu} c_{m \mu} \partial_{\rho_{1}}\phi^{J_{1}}_{m}\partial_{\rho_{2}}\phi^{J_{2}}_{m }\partial_{\rho_{3}}\phi^{J_{3}}_{m }\phi^{L}_{m }}{|c_{m}|(\sqrt{\phi^{I}_{m } \phi_{m I}})^{5}} Sp[\Gamma^{\rho}\Gamma^{\nu} \Gamma^{\rho_{1}}\Gamma^{\rho_{2}}\Gamma^{\rho_{3}}  \Gamma^{\mu}  \Gamma_{7}\Gamma_{J_{1}} \Gamma_{J_{2}}\Gamma_{J_{3}}\Gamma_{L}\Gamma_{I} ] 	
\end{equation}
To get the nonzero result, the first six Gamma matrices should multiply to $1$, while the last five Gamma matrices should cover $\Gamma_{6}\cdots \Gamma_{10}$. The trace then becomes $Sp[\Gamma_{0}\cdots \Gamma_{10}] \sim \epsilon_{0\cdots 10}$. $\rho_{1}$, $\rho_{2}$ and $\rho_{3}$ must be different, so $\left\{\rho, \nu, \mu \right\}=\left\{\rho_{1}, \rho_{2}, \rho_{3} \right\}$. One can similarly expand $f_{m\rho\nu}$ as $f_{m\rho\nu} = h_{m \rho\nu \sigma} c^{\sigma}_{m}$. $f_{m\rho\nu} c_{m \mu} = h_{m \rho\nu \sigma} c^{\sigma}_{m} c_{m \mu}$. As is mentioned before, $c_{m}$ is not fixed but should also be integrated in the path integral. The orientation of the selfdual string in $6d$ spacetime is arbitrary. Summing over all possible directions, $\sum c^{\sigma}_{m} c_{m \mu} = |c_{m}|^{2} g^{\sigma}_{\mu}$, so $\sum f_{m\rho\nu} c_{m \mu} = h_{m \rho\nu \sigma} \sum c^{\sigma}_{m} c_{m \mu} = |c_{m}|^{2} h_{m \rho\nu \mu} $. (\ref{32}) then becomes
\begin{equation}\label{34}
	\frac{\delta L_{1}}{\delta  \phi^{I}_{m }(x)} \propto \frac{ |c_{m}| (*h_{m})_{\rho\nu\mu} \partial_{\rho_{1}}\phi^{J_{1}}_{m }\partial_{\rho_{2}}\phi^{J_{2}}_{m }\partial_{\rho_{3}}\phi^{J_{3}}_{m} \phi^{L}_{m }}{(\sqrt{\phi^{I}_{m } \phi_{m I}})^{5}}\epsilon^{\rho\nu\mu\rho_{1}\rho_{2}\rho_{3}}\epsilon_{J_{1}J_{2}J_{3}LI}. 
\end{equation}
Compared with (\ref{33}), (\ref{34}) contains $*$ which is resulted from $\Gamma_{7}$ inside the trace. This is not quite satisfactory, but luckily, since $*h=h$, (\ref{33}) and (\ref{34}) still coincide up to a $|c_{m}|$ factor.

Notice that to get $L_{1}$ which is close to $L'_{1}$, $d=5$ is quite crucial. For an ordinary $6d$ theory without the constraint, the denominator is $(\sqrt{\phi^{I}_{m \mu} \phi^{\mu}_{m I}})^{2n-4}$, so one cannot get $L'_{1}$ no matter which $n$ is taken. In other words, to get the WZ term $H_{3} \wedge A_{3}$, the basic degrees of freedom should be the $1d$ object with $5d$ momentum other than the $0d$ object with $6d$ momentum. For $1/2$ BPS selfdual strings, the momentum is along the transverse direction, since the momentum along the longitudinal direction may reduce the selfdual string to a $1/4$ BPS state \cite{ccc}. So the constraint in \cite{Z} may indicate that the selfdual strings involved in equations are $1/2$ BPS states. The WZ term $H_{3} \wedge A_{3}$ is generated by $1/2$ BPS selfdual strings.

In \cite{Ber}, the anomaly of the $1/2$ BPS selfdual strings coupling with tensor multiplets was discussed in analogy with the anomaly of M5 branes coupling with supergravity. In the generic Coulomb branch, there are $N(N-1)$ selfdual strings acting as the sources for the relative 3-forms $h_{ij}$. $H_{3} \wedge A_{3}$ is exactly the term in the bulk cancelling the selfdual string normal bundle anomaly \cite{Ber}. When reduced to $5d$, $H_{3} \wedge A_{3}$ becomes $F_{2} \wedge A_{3}$, which is the term needed to cancel the normal bundle anomaly of the magnetic monostring \cite{Ber}. On the other hand, $F_{2} \wedge A_{3}$ could be obtained in $5d$ SYM theory by the integration out of the massive fields which are the quantization of the electric dual of the monostring. The dual of the selfdual string is itself, so we may expect that $H_{3} \wedge A_{3}$ could be derived by the integration out of the fields which are the quantization of the selfdual string.

The next problem is to get the WZ term corresponding to $A_{3}\wedge F_{4} \sim \sigma_{3}\wedge \omega_{4}$, which is far more difficult. The field theory calculation gives $\delta \Gamma_{WZ}/\delta \phi^{I}$, the counterpart of the Lorentz force on supergravity side. Consider a M5 brane in a background field $\hat{F}_{4}$, the action contains the term $S = - \int_{W_{6}} A_{6}$, $d A_{6} = *\hat{F}_{4}+ A_{3}\wedge \hat{F}_{4}/2$. 
\begin{eqnarray}\label{2wq}
&& \frac{\delta S}{\delta \phi^{I}}	=  - \frac{1}{6!}\epsilon^{\rho_{1}\rho_{2}\rho_{3}\rho_{4}\rho_{5}\rho_{6}} \partial_{\rho_{1}} Y^{n_{1}}\partial_{\rho_{2}}  Y^{n_{2}}\partial_{\rho_{3}}  Y^{n_{3}}\partial_{\rho_{4}}  Y^{n_{4}}\partial_{\rho_{5}}  Y^{n_{5}}\partial_{\rho_{6}}  Y^{n_{6}}(d A_{6})_{In_{1}n_{2}n_{3}n_{4}n_{5}n_{6}}
 \nonumber \\
&=& - \frac{1}{6!}\epsilon^{\rho_{1}\rho_{2}\rho_{3}\rho_{4}\rho_{5}\rho_{6}} \partial_{\rho_{1}} Y^{n_{1}}\partial_{\rho_{2}}  Y^{n_{2}}\partial_{\rho_{3}}  Y^{n_{3}}\partial_{\rho_{4}}  Y^{n_{4}}\partial_{\rho_{5}}  Y^{n_{5}}\partial_{\rho_{6}}  Y^{n_{6}} (*\hat{F}_{4}+ \frac{1}{2}A_{3}\wedge \hat{F}_{4})_{In_{1}n_{2}n_{3}n_{4}n_{5}n_{6}}
\nonumber \\
&=& f_{6I}+g_{6I}. 
\end{eqnarray} 
$Y^{n_{i}}$ are embedding coordinates. $n_{i} = 0\cdots 10$, $Y^{I} = \phi^{I}$. $f_{6I}$ and $g_{6I}$ are forces related with the magnetic-magnetic interaction and the electric-magnetic interaction respectively. $g_{6I}$ is the Lorentz force derived from the WZ term. If $db_{2}$ is also taken into account,
\begin{equation}\label{wq2}
	g_{6I} = \frac{1}{6!}\epsilon^{\rho_{1}\rho_{2}\rho_{3}\rho_{4}\rho_{5}\rho_{6}} \partial_{\rho_{1}} Y^{n_{1}}\partial_{\rho_{2}}  Y^{n_{2}}\partial_{\rho_{3}}  Y^{n_{3}}\partial_{\rho_{4}}  Y^{n_{4}}\partial_{\rho_{5}}  Y^{n_{5}}\partial_{\rho_{6}}  Y^{n_{6}} ( \frac{1}{2}h_{3}\wedge \hat{F}_{4})_{In_{1}n_{2}n_{3}n_{4}n_{5}n_{6}}.
\end{equation}

For the WZ term in (\ref{l}),\footnote{Here, for simplicity, the last line in (\ref{44}) is denoted as the differential form, but it should be more accurately written in the form like that in (\ref{2wq}) and (\ref{wq2}). }  
\begin{eqnarray}\label{44}
\frac{\delta \int_{W_{7i}}\sigma_{3ji}\wedge \omega_{4ki} }{\delta \phi^{I}_{i}(x)}	&=& -\int_{W_{7i}} [  \frac{\delta \sigma_{3ji}}{\delta \phi^{I}_{ji}(x) }\wedge \omega_{4ki}+ \sigma_{3ji}\wedge \frac{\delta  \omega_{4ki}}{\delta \phi^{I}_{ki}(x)}]
 \nonumber \\
&=& (\sigma_{3ji} \wedge F_{3kiI})(x) - (\omega_{4ji} \wedge F_{3kiI} + \omega_{4ki} \wedge F_{3jiI})(x). 
\end{eqnarray} 
$\partial W_{7i} = W_{6i}$, $x \in W_{6i}$.  
\begin{equation}
	F_{3kiI} = \alpha \; \frac{\epsilon_{ILJ_{1}J_{2}J_{3}}\phi^{L}_{ki}d \phi^{J_{1}}_{ki}\wedge d \phi^{J_{2}}_{ki} \wedge d \phi^{J_{3}}_{ki}}{|\phi_{ki}|^{5}}
\end{equation}
up to a total derivative. $\alpha$ is a constant. Except for the 6-form, (\ref{44}) also contains the 7-form because $\sigma_{3}\wedge \omega_{4}$ alone is not closed.
\begin{equation}\label{40}
	\frac{\delta \int_{W_{7i}}(\sigma_{3ji}\wedge \omega_{4ki} + \sigma_{3ki}\wedge \omega_{4ji})}{\delta \phi^{I}_{i}(x)} = (\sigma_{3ji} \wedge F_{3kiI} +  \sigma_{3ki} \wedge F_{3jiI})(x)-2(\omega_{4ji} \wedge F_{3kiI} + \omega_{4ki} \wedge F_{3jiI})(x), 
\end{equation}
\begin{eqnarray}\label{41}
\frac{\delta \int_{W_{7}} \Omega_{ijk}}{\delta \phi^{I}_{i}(x)}	& \propto &   (\sigma_{3ji}\wedge F_{3kiI} +  \sigma_{3ki} \wedge F_{3jiI} - \sigma_{3kj} \wedge F_{3ijI}- \sigma_{3jk} \wedge F_{3ikI} )(x)
 \nonumber \\
&&  - 2( \omega_{4ji}\wedge F_{3kiI} + \omega_{4ki}\wedge F_{3jiI} - \omega_{4kj} \wedge F_{3ijI} - \omega_{4jk}  \wedge F_{3ikI})(x).
\end{eqnarray}

We need to get the 6-form of (\ref{41}) from the field theory calculation. The first thing is to find an explicit expression for $\sigma_{3}$. Consider the $5d$ transverse space with coordinate $\phi^{a_{i}}$, $a_{i} = 6\cdots 10$, 
\begin{equation}
\omega_{4} = \epsilon_{a_{1}a_{2}a_{3}a_{4} a_{5}}\frac{1}{|\phi|^{5}}\phi^{a_{1}}d\phi^{a_{2}}\wedge d\phi^{a_{3}}\wedge d\phi^{a_{4}} \wedge d\phi^{a_{5}} = *\hat{\phi}. 	
\end{equation}
$\sigma_{3}$ could be constructed in analogy with the gauge field describing the magnetic monopole in $3d$ space. Select an arbitrary vector $v$, $v \cdot \phi= |v||\phi| \cos \theta$, 
\begin{eqnarray}\label{37}
\sigma_{3}	&=& \frac{(\cos^{3} \theta -3\cos \theta+2 )*(v\wedge \phi)}{3 \sin^{4} \theta |v||\phi|} \nonumber \\
&=& \frac{(v \cdot \phi)^{3} -3 (v \cdot \phi)|v|^{2}|\phi|^{2}+2|v|^{3}|\phi|^{3} }{3 \sin^{4} \theta |v|^{4}|\phi|^{7}} \epsilon_{a_{1}a_{2}a_{3}a_{4} a_{5}}v^{a_{1}}\phi^{a_{2}}  d\phi^{a_{3}}\wedge d\phi^{a_{4}} \wedge d\phi^{a_{5}}. 
\end{eqnarray} 
$\sigma_{3}$ is singular on a ray $OV$ starting from the origin and extending in $-v$ direction. In $11d$ spacetime, $OV \times W_{6} = W_{7}$. $W_{7}$ is the Dirac brane similar to the Dirac string \cite{a1, a2}. Replace $v$ by $-v$, we get another $\sigma_{3}$. Take the average of these two $\sigma_{3}$'s, the last term in (\ref{37}) could be dropped. Of course, in this case, the singularity exists in a straight line. One may have  
\begin{equation}\label{110}
	\sigma_{3ji} = \frac{(v \cdot \phi_{ji})^{3} -3 (v \cdot \phi_{ji})|v|^{2}|\phi_{ji}|^{2}}{3 \sin^{4} \theta |v|^{4}|\phi_{ji}|^{7}} \epsilon_{a_{1}a_{2}a_{3}a_{4} a_{5}} v^{a_{1}}\phi_{ji}^{a_{2}} \; d\phi_{ji}^{a_{3}}\wedge d\phi_{ji}^{a_{4}} \wedge d\phi_{ji}^{a_{5}} + d\chi_{2ji},  
\end{equation}
$d \sigma_{3ji}= \omega_{4ji}$.

$\sigma_{3ji}$ is only determined up to an exact form $d\chi_{2ji}$. One may want to take the first term of (\ref{110}) as $\sigma_{3ji}$ for simplicity, but there is a problem. The only input on field theory side is the vacuum expectation value $( \phi_{1}, \cdots, \phi_{N} )$ on $W_{6}$. As a result, we have to construct the 6-form from $(\phi_{ij}, \phi_{jk}, \phi_{ki})$. The first term of $\sigma_{3ji}$ in (\ref{110}) contains a constant vector $v$, which has no relevance with the vacuum expectation value. We may try to replace $v$ by vectors constructed from $( \phi_{1}, \cdots, \phi_{N} )$. For example, for the 6-form $\sigma_{3ji} \wedge F_{3kiI}$ living on $W_{6}$, 
\begin{eqnarray}\label{1011}
 \sigma_{3ji} \wedge F_{3kiI}   &=&  \alpha \;  \frac{(v \cdot \phi_{ji})^{3} -3 (v \cdot \phi_{ji})|v|^{2}|\phi_{ji}|^{2}}{3 \sin^{4}\theta |v|^{4} |\phi_{ji}|^{7}|\phi_{ki}|^{5}} \epsilon_{I b_{1}b_{2}b_{3}b_{4}} \epsilon_{L a_{1}a_{2}a_{3}a_{4}}  v^{L}\phi_{ji}^{a_{1}}
 \nonumber \\
&& \phi^{b_{1}}_{ki}
\; d\phi_{ji}^{a_{2}}\wedge d\phi_{ji}^{a_{3}} \wedge d\phi_{ji}^{a_{4}}\wedge d \phi^{b_{2}}_{ki}\wedge d \phi^{b_{3}}_{ki} \wedge d \phi^{b_{4}}_{ki}. 
\end{eqnarray} 
$v$ may be replaced by $\phi_{ki}$, but then the extra terms should be added because $\phi_{ki}$ may not be constant.

Notice that the epsilon symbols involved in $H_{3}\wedge F_{3}$ and the Lorenz force in SYM theories are $\epsilon_{M_{1}M_{2}\cdots M_{11}}$ and $\epsilon_{N_{1}N_{2}\cdots N_{10}}$ which are equal to the traces of the Gamma matrices product, thus could be derived via the 1-loop fermion integration. On the other hand, $\sigma_{3} \wedge F_{3}$ contains $\epsilon_{a_{1}\cdots a_{5}}\epsilon_{b_{1}\cdots b_{5}}\epsilon_{\mu_{1} \cdots \mu_{6}}$ with $a_{i}, b_{i} = 6\cdots 10$, $\mu_{i}=0\cdots 5$ that cannot be identified with the single trace. As a result, a 1-loop calculation cannot produce $A_{3}\wedge F_{4}$. Maybe a refined integration is needed. For example, since each fermion loop gives a single epsilon symbol, one may consider two fermion loops, one for $\phi_{ji}$ and the other for $\phi_{ki}$, giving rise to $\epsilon_{a_{1}\cdots a_{5}}\epsilon_{\mu_{1} \cdots \mu_{3}\nu_{1} \cdots \nu_{3}}$ and $\epsilon_{b_{1}\cdots b_{5}}\epsilon_{\mu_{4} \cdots \mu_{6}\nu_{4} \cdots \nu_{6}}$ respectively. With a suitable contraction, $\epsilon_{a_{1}\cdots a_{5}}\epsilon_{b_{1}\cdots b_{5}}\epsilon_{\mu_{1} \cdots \mu_{3}\nu_{1} \cdots \nu_{3}}\epsilon_{\mu_{4} \cdots \mu_{6}}\!^{\nu_{1} \cdots \nu_{3}}$ may be produced, from which, one can at most get $*A_{3}\wedge F_{3}$ other than $A_{3}\wedge F_{3}$. This is acceptable if the self duality condition is imposed. If we take $\psi_{ij}$ as the fundamental fields, the two loop fermion integration seems indicate that the anomaly is produced by the bound state of $\psi_{ji}$ and $\psi_{ki}$, which are again three index objects. Given that in standard field theory, the only way to get the epsilon symbols is through the fermion loop integration, one may also expect that the $6d$ $(2,0)$ theory may not be the ordinary QFT thus cannot be analyzed in this way. No matter in which case, even if we could get the correct epsilon symbol, we still need to ensure that each term takes the form of $\sigma_{3ji}\wedge \omega_{4ki}$ with two roots involved. One possibility is to start from the three indices fundamental fields at the beginning. Or one can take two indices fields as fundamental, but there must be a justification for why $\psi_{ji}$ and $\psi_{ki}$ appear simultaneously in the integration.

\section{Conclusion}

In this paper, we discussed the WZ term in the low energy effective action of the $6d$ $(2, 0)$ field theory in the generic Coulomb branch. As a topological term, WZ term does not depend on the metric nor the coupling, so it is protected. For such terms, the supergravity calculation and the field theory calculation will give the same result. There is no available $6d$ $(2,0)$ field theory at present, so we will first calculate the WZ term on supergravity side. We then show that the obtained WZ term could indeed compensate the anomaly deficit, as is required by the anomaly matching condition, thus should appear in the low energy effective action.

For SYM theory in a generic Coulomb branch, each WZ term involves one root $e_{i}-e_{j}$, which is consistent with the fact that the supergravity interaction is produced by the integrating out of massive strings connecting the $i_{th}$ and the $j_{th}$ D branes. On the other hand, for M5 branes, each WZ term involves two roots $e_{i}-e_{j}$ and $e_{k}-e_{j}$. One may expect that such kind of triple interaction may be generated by the integrating out of the massive objects carrying $(i, j, k)$ indices. A natural candidate is the string junction with tension $(|\phi_{ij}|,|\phi_{jk}|,|\phi_{ki}|)$ proposed in \cite{K, L}.

The $6d$ $(2,0)$ theory may have a mathematical structure which is different from the ordinal QFT, or it could just be constructed as a normal quantum field theory. In the latter case, the WZ term could be calculated by a standard 1-loop integration. The input from the field theory is the Dirac operator on the given background scalar fields and tensor fields. The algebra structure is involved. For $SU (N) $ SYM theories, the 2-algebra gives $(N^{2}-N)/2$ terms characterized by roots $e_{i}-e_{j}$. The Hopf-Wess-Zumino term for $6d$ $(2,0)$ theory is composed by $(N^{3}-N)/3$ terms with roots $(e_{i}-e_{j},e_{j}-e_{k},e_{k}-e_{i})$, so it may indicate a new algebra structure. The 3-algebra may be a natural candidate, but it has no finite dimensional Euclidean representation. We calculate the WZ term for the 3-algebra valued (2, 0) tensor multiplet theory proposed in \cite {Z}. The $H_{3}\wedge A_{3}$ part of the WZ term could be obtained. However, the $A_{3}\wedge F_{4}$ part, which is responsible for the anomaly compensating, cannot be produced by the 1-loop fermion integration, indicating that some key ingredient is still missing.

\bigskip
\bigskip

{\bf Acknowledgments:}
We are grateful to Ken Intriligator for helpful comment. The work is supported in part by the Mitchell-Heep Chair in High Energy Physics and by the DOE grant DE-FG03-95-Er-40917.

\appendix

\section{The on-shell action for brane-gravity coupled system}

Consider the $(d-1)$-brane couples with the supergravity fields. The action is \cite{B}
\begin{equation}
	S = S_{brane} + S_{gravity}
\end{equation}
\begin{eqnarray}\label{400}
S_{brane} & =&  T_{d} \int d^{d}\xi	[ -\frac{1}{2} \sqrt{-\gamma}\gamma^{ij}\partial_{i} X^{M} \partial_{j} X^{N} g_{MN} e^{\alpha (d)\phi/d} + \frac{d-2}{2}\sqrt{-\gamma}
 \nonumber \\
& & - \frac{1}{d!}\epsilon^{i_{1}i_{2} \cdots i_{d}} \partial_{i_{1}} X^{M_{1}} \partial_{i_{2}} X^{M_{2}} \cdots \partial_{i_{d}} X^{M_{d}} A_{M_{1}M_{2}\cdots M_{d}}] \nonumber \\ & =& S_{1}+ S_{2} + S_{3}
\end{eqnarray} 
\begin{equation}
	S_{gravity} = \frac{1}{2 \kappa^{2}} \int d^{D} x \sqrt{-g} [ R - \frac{1}{2} (\partial \phi)^{2} - \frac{1}{2(d+1)!} e^{- \alpha (d)\phi} F^{2}_{d+1}] 
\end{equation}
The action of this form is valid for both electric and magnetic branes, while for magnetic branes, just let $\alpha(d)\rightarrow -\alpha(d)$. Variation with respect to $g_{MN}$, $A_{M_{1}M_{2}\cdots M_{d}}$, and $\gamma_{ij}$ gives 
\begin{eqnarray}
T^{MN}_{brane} & =&  - T_{d} \int d^{d}\xi \sqrt{-\gamma}\gamma^{ij}\partial_{i} X^{M} \partial_{j} X^{N}  e^{\alpha (d)\phi/d} \frac{\delta^{D} (x-X)}{\sqrt{-g}} 
 \nonumber \\
 & =& \frac{1}{\kappa^{2}}  [ R^{MN} - \frac{1}{2}g^{MN} R - \frac{1}{2} (\partial^{M} \phi \partial^{N} \phi - \frac{1}{2}g^{MN}(\partial \phi)^{2}  ) \nonumber \\
 & & - \frac{1}{2 d!} ( F^M\,_{M_1 \cdots M_d}  F^{N M_{1}\cdots M_{d}} - \frac{1}{2(d+1)} g^{MN} F^{2}) e^{- \alpha (d)\phi}] 
\end{eqnarray} 
\begin{eqnarray}\label{600}
 J^{M_1 \cdots M_d}_{brane} & =&  T_{d} \int d^{d}\xi \, \epsilon^{i_{1}i_{2} \cdots i_{d}} \partial_{i_{1}} X^{M_{1}} \partial_{i_{2}} X^{M_{2}} \cdots \partial_{i_{d}} X^{M_{d}} \frac{\delta^{D} (x-X)}{\sqrt{-g}}
\nonumber \\
 & =& \frac{1}{2\kappa^{2}\sqrt{-g}} \partial_{M} (\sqrt{-g}e^{- \alpha (d)\phi} F^{M M_{1}\cdots M_{d}} )
\end{eqnarray} 
and
\begin{equation}\label{300}
	\gamma_{ij} = \partial_{i} X^{M} \partial_{j} X^{N} g_{MN} e^{\alpha (d)\phi/d}. 
\end{equation}
Then 
\begin{eqnarray}
S_{1} & =&  \frac{1}{2} \int d^{D} x \sqrt{-g} \, T^{MN}_{brane} \, g_{MN}
\nonumber \\
 & =& \frac{1}{2 \kappa^{2}}  \int d^{D} x \sqrt{-g}  [ (1- \frac{D}{2} )(R- \frac{1}{2}(\partial \phi)^{2}) - \frac{1}{2 d!} ( 1 - \frac{D}{2(d+1)})  F^{2} e^{- \alpha (d)\phi}]	
\end{eqnarray} 
Plug (\ref{300}) into (\ref{400}), we get
\begin{equation}\label{f}
	S_{1}+S_{2}= \frac{2}{d} S_{1}
\end{equation}
From (\ref{600}),
\begin{eqnarray}
S_{3} & =&  - \frac{1}{d!} \int d^{D} x   \sqrt{-g} \, J^{M_1 \cdots M_d}_{brane} A_{M_{1}M_{2}\cdots M_{d}}
\nonumber \\ 
& =& \frac{1}{2 \kappa^{2}} \int d^{D} x  \frac{1}{(d+1)!}  \sqrt{-g}e^{- \alpha (d)\phi} F^{2}\nonumber \\ 
&&- \frac{1}{d!} \partial_{M}  (\sqrt{-g}e^{- \alpha (d)\phi} A_{M_{1}M_{2}\cdots M_{d}} F^{M M_{1}\cdots M_{d}}) 
\end{eqnarray} 
As a result, 
\begin{eqnarray}
S_{brane} & =&  S_{1}+ S_{2} + S_{3}
\nonumber \\
 & =& \frac{2-D}{d}  \frac{1}{2 \kappa^{2}} \int d^{D} x \sqrt{-g} [ R - \frac{1}{2} (\partial \phi)^{2} - \frac{1}{2(d+1)!} e^{- \alpha (d)\phi} F^{2}_{d+1}] 
\nonumber \\
 && - \frac{1}{2 \kappa^{2}} \int d^{D} x  \frac{1}{d!} \partial_{M}  (\sqrt{-g}e^{- \alpha (d)\phi} A_{M_{1}M_{2}\cdots M_{d}} F^{M M_{1}\cdots M_{d}})  \nonumber \\
 & =& \frac{2-D}{d}  S_{gravity}+S_{boundary}
\end{eqnarray} 
For $d<3$, or equivalently, for purely electric branes, $S_{boundary}=0$. If $M_{D}$ has no boundary, $S_{boundary}$ could also be dropped. Then for the given brane configuration, if $S$ is on-shell with respect to supergravity, we have 
\begin{equation}\label{700}
S_{brane} : S_{gravity} : S = (D-2) : (-d) : (D-2-d),	
\end{equation}
where
\begin{eqnarray}
S_{brane} & =&  T_{d} \int d^{d}\xi	[- \sqrt{-\det (\partial_{i} X^{M} \partial_{j} X^{N} g_{MN} e^{\alpha (d)\phi/d})}
 \nonumber \\
& & - \frac{1}{d!}\epsilon^{i_{1}i_{2} \cdots i_{d}} \partial_{i_{1}} X^{M_{1}} \partial_{i_{2}} X^{M_{2}} \cdots \partial_{i_{d}} X^{M_{d}} A_{M_{1}M_{2}\cdots M_{d}}] 
\end{eqnarray} 
The extension to multi-brane configurations is straightforward, and (\ref{700}) still holds. When the dimensions of the branes are different, the exact proportional relation is not valid anymore. Besides, when $F_{ij}$ does not vanish, i.e. the $(p-1) $-brane carries $p-1-2n$ charge, (\ref{700}) does not hold. Naively, when $D = 11$, $d=6$, neglecting the boundary term, $S_{brane} : S_{gravity} : S = 3 : (-2) : 1$\footnote{Note that for D3, M2, M5 branes, $S_{brane}/S$ equals to $2$, $3/2$, and $3$, while the degrees of freedom on these branes scale as $N^{2}$, $N^{3/2}$, and $N^{3}$. This is not the coincidence. Suppose the degrees of freedom on N branes scale as $N^{\alpha}$. Also suppose that in $S_{brane}$, there is a term $T$ has the $N^{\alpha}$ scaling. Consider $N+1$ branes with large $N$, when the symmetry is broken from $SU(N+1)$ to $SU(N)\times U(1)$, $(N+1)^{\alpha}-N^{\alpha}\sim \alpha N^{\alpha-1}$ number of $T$ will enter into $S_{brane}$. On the other hand, the effective action of the system could be approximated as the action of a single brane on the background generated by the rest $N$ branes. In $S_{eff}$, one may get $N^{\alpha-1}T$. Obviously, $T$ in $S_{brane}$ and $T$ in $S_{eff}$ differ by a $\alpha$ factor. For $D3$, $T$ is $\int F_{5}$, for $M5$, $T$ is $\int A_{3}\wedge F_{4}$, while for $M2$, $T$ is obscure.}. However, $S$ here is not exactly the action for M5 branes coupling with supergravity. In the following, we will use (\ref{a}) as the action to get the same conclusion.

Now, consider 
\begin{equation}
S=S_{g}+S_{M5}	
\end{equation}
with 
\begin{equation}
S_{g} = \frac{1}{2 \kappa^{2}}\int_{M_{11}} *R-\frac{1}{2} *\hat{F}_{4}\wedge \hat{F}_{4} - \frac{1}{6} F_{4}\wedge F_{4} \wedge A_{3}	
\end{equation}
\begin{eqnarray} 
S_{M5} &=&  - T_{5}\int_{W_{6}} d^{6}\xi \sqrt{-\det(g_{\mu \nu}+ (i_{v_{1}} \tilde{*} h_{3})_{\mu \nu})} + \frac{1}{2} v_{1}\wedge h_{3} \wedge \tilde{*} (v_{1}\wedge \tilde{*} h_{3})
 \nonumber \\
&+& \frac{T_{5}}{2} \int_{W_{6}} db_{2} \wedge A_{3} + \frac{T_{5}}{2} \int_{W_{7}} A_{3} \wedge F_{4}
\end{eqnarray} 
The field equations are \cite{a1}
\begin{equation}\label{d}
	T^{MN}_{M5} = \frac{1}{\kappa^{2}}\left[ R^{MN} - \frac{1}{2}g^{MN} R - \frac{1}{12}(\hat{F}^{M}_{4}\!_{PQL}\hat{F}^{NPQL}_{4}-\frac{1}{8}g^{MN}\hat{F}^{2}_{4}) \right]
\end{equation}
\begin{equation}\label{e}
	d* \hat{F}_{4}+ \frac{1}{2}F_{4} \wedge F_{4} = -2\kappa^{2} T_{5}( -A_{3}\wedge *J_{6} +F_{4}\wedge *G_{7} )
\end{equation}
\begin{equation}
	d \hat{F}_{4} = 2\kappa^{2}T_{5} *J_{6}, 
\end{equation}
The vacuum expectation value of $b_{2}$ are taken to be zero, otherwise, (\ref{700}) does not hold. From (\ref{d}), 
\begin{equation}
\frac{1}{2} \int_{W_{6}} d^{6} \xi \sqrt{-g} \, T^{MN}_{M5} \, g_{MN}	= \frac{1}{2\kappa^{2}}\int_{M_{11}} -
\frac{9}{2} *R +\frac{3}{4} *\hat{F}_{4} \wedge \hat{F}_{4}
\end{equation}
From (\ref{e}), 
\begin{equation}\label{g}
	\frac{T_{5}}{2} \int_{W_{7}} A_{3} \wedge F_{4} = \frac{1}{2\kappa^{2}}\int_{M_{11}}\frac{1}{2}*\hat{F}_{4} \wedge F_{4} + \frac{1}{4} F_{4}\wedge F_{4}\wedge A_{3}-\frac{1}{4 \kappa^{2}} \int_{\partial M_{11}} A_{3} \wedge * \hat{F}_{4}
\end{equation}
We will still use the general relation (\ref{f}) for M5 branes. The Nambu-Goto action for M5 branes is more involved than that for D branes or M2 branes, so the correction may exist, but that will not bring too many problems, since our main concern is the WZ term.
\begin{equation}
	S_{M5} = \frac{1}{2\kappa^{2}}\int_{M_{11}} -
\frac{3}{2} *R +\frac{3}{4} *\hat{F}_{4} \wedge \hat{F}_{4} + \frac{1}{4} F_{4}\wedge F_{4}\wedge A_{3} + \frac{T_{_{5}}}{2}\int _{W_{7}} * \hat{F}_{4}-\frac{1}{4 \kappa^{2}} \int_{\partial M_{11}} A_{3} \wedge * \hat{F}_{4}
\end{equation}
\begin{equation}
	S = \frac{1}{2\kappa^{2}}\int_{M_{11}} -
\frac{1}{2} *R +\frac{1}{4} *\hat{F}_{4} \wedge \hat{F}_{4} + \frac{1}{12} F_{4}\wedge F_{4}\wedge A_{3} + \frac{T_{_{5}}}{2}\int _{W_{7}} * \hat{F}_{4}-\frac{1}{4 \kappa^{2}} \int_{\partial M_{11}} A_{3} \wedge * \hat{F}_{4}
\end{equation}
Because of the last two terms, the exact proportional relation does not hold. We are interested with (\ref{g}), which could be rewritten as
\begin{equation}
	\frac{T_{5}}{2} \int_{W_{7}} A_{3} \wedge F_{4} = \frac{1}{2\kappa^{2}}\int_{M_{11}} \frac{1}{4} F_{4}\wedge F_{4}\wedge A_{3} + \frac{1}{2} A_{3} \wedge d* \hat{F}_{4}
\end{equation}
For magnetic field, $d* \hat{F}_{4}=0$, the last term vanishes, so the Hopf-Wess-Zumino term is only related with $F_{4}\wedge F_{4}\wedge A_{3}$, for which, the $1/3$ factor appears.

Similarly, for D3 branes, one may expect that the WZ term should be one half of the corresponding term in the D3 brane action. This is indeed the case, and it is just this rescaled term that is obtained from the 1-loop integration and compensates the anomaly deficit.


\begin{thebibliography}{10}



\bibitem{1}
E.\ Witten, ``Global aspects of current algebra'', Nucl. Phys. B 223 (1983) 422.


\bibitem{2}
A.\ V. Manohar, ``Wess-Zumino Terms in Supersymmetric Gauge Theories'', Phys. Rev. Lett. 81 (1998) 1558, hep-th/9805144.

\bibitem{3}
G.\ 't Hooft, Recent Developments in Gauge Theories, eds. G.\ 't Hooft et. al., Ple Press, NY, 1980.


\bibitem{A}
K.\ A. Intriligator, ``Anomaly matching and a Hopf-Wess-Zumino term in 6d, N=(2,0)
field theories'', \NP {\bf B581} 257 (2000), hep-th/0001205.

\bibitem{D}
C.\ Boulahouache and G.\ Thompson, ``One loop effects in various dimensions and D-branes'', Int. J. Mod. Phys. A13, 5409 (1998), hep-th/9801083.


\bibitem{E}
A.\ A. Tseytlin and K.\ Zarembo, ``Magnetic interactions of D-branes and Wess-Zumino terms in super Yang-Mills effective actions'', Phys. Lett. B 474, 95 (2000), hep-th/9911246.

\bibitem{J}
O.\ Ganor and L.\ Motl, ``Equations of the $(2,0)$ theory and knitted fivebranes'', JHEP 05, 009 (1998), hep-th/9803108.




\bibitem{K}
K.\ M. Lee and H.\ U. Yee, ``BPS string webs in the 6-dim (2,0) theories'', JHEP 0703 (2007) 057, hep-th/0606150.




\bibitem{L}
S.\ Bolognesi and K.\ M. Lee, ``$1/4$ BPS string junctions and $N^{3}$ problem in 6-dim (2,0) superconformal theories'', [arXiv:1105.5073 [hep-th]].

\bibitem{F}
D.\ V. Belyaev and I.\ B. Samsonov, ``Wess-Zumino term in the N=4 SYM effective action revisited'', JHEP 04, 112 (2011), [arXiv:1103.5070 [hep-th]]. 


\bibitem{Z}
N. Lambert and C. Papageorgakis, ``Nonabelian (2,0) tensor multiplets and 3-algebras'', JHEP 08 (2010) 083, [arXiv:1007.2982 [hep-th]].

\bibitem{a1}
M. S. Bremer, ``Vacua and p-branes in Maximal Supergravities''.

\bibitem{a11}
D. P. Sorokin, ``Coupling of M-branes in M-theory'', Talk given at 6th International Symposium on Particles, Strings and
Cosmology (PASCOS 98), Boston, MA, 22-27 Mar 1998. Published in Boston 1998, Particles, strings and cosmology, 697-701, hep-th/9806175.

\bibitem{a2}
S. P. de Alwis, ``Coupling of branes and normalization of effective actions in string/M-theory'', Phys. Rev. D56 (1997) 7963, hep-th/9705139.

\bibitem{C}
Y.\ Okawa, T.\ Yoneya, ``Multi-Body Interactions of D-Particles in Supergravity and Matrix Theory'', Nucl. Phys. B538 (1999) 67-99, hep-th/9806108.





\bibitem{C1a}
D. Freed, J. A. Harvey, R. Minasian, and G. W. Moore, ``Gravitational anomaly
cancellation for M-theory fivebranes'', Adv. Theor. Math. Phys. 2 (1998) 601-618, hep-th/9803205.   

\bibitem{C1b}
J. A. Harvey, R. Minasian, and G. W. Moore, ``Non-abelian tensor-multiplet anomalies'', JHEP 09, 004 (1998), hep-th/9808060.  

\bibitem{C2a} 
L Alvarez-Gaume and E Witten, ``Gravitational anomalies'', Nucl. Phys. B 234 (1983) 269.

\bibitem{C2b} 
E. Witten, ``Five-brane effective action in M-theory'', J. Geom. Phys. 22 (1997) 103, hep-th/9610234.

\bibitem{G}
S.\ Ferrara, R.\ R. Khuri and R.\ Minasian, ``M Theory on a Calabi-Yau Manifold'', Phys. Lett. B375 (1996) 81, hep-th/9602102. 
 
\bibitem{H}
J.\ Maldacena, A.\ Strominger and E.\ Witten, ``Black Hole Entropy in M theory'', JHEP 9712, 002 (1997), hep-th/9711053. 

\bibitem{III}
I. R. Klebanov and A. A. Tseytlin, ``Intersecting M-branes as four-dimensional black holes'', Nucl. Phys. B 475, 179 (1996), hep-th/9604166.

\bibitem{I}
D.\ Berenstein and R.\ Leigh, ``String junctions and bound states of intersecting branes'', \PR {\bf D60} (1999) 026005, hep-th/9812142.






\bibitem{M}
A.\ A. Tseytlin, ``$R^{4}$ terms in $11$ dimensions and conformal anomaly of $(2, 0)$
theory'', Nucl. Phys. B584 (2000) 233-250, hep-th/0005072.


\bibitem{N}
O.\ Bergman, ``Three Pronged Strings and 1/4 BPS States in N = 4 Super Yang-Mills Theory'', Nucl. Phys. B525 (1998) 104, hep-th/9712211.




\bibitem{P}
J.\ M. Maldacena and C.\ Nunez, ``Supergravity description of field theories on curved manifolds and a no go theorem'', Int. J. Mod. Phys. A 16 (2001) 822, hep-th/0007018.


\bibitem{Q}
D.\ Gaiotto, ``N = 2 Dualities'', [arXiv:0904.2715 [hep-th]].


\bibitem{R}
D.\ Gaiotto, J.\ Maldacena, ``The Gravity duals of N=2 superconformal field theories'', [arXiv:0904.4466 [hep-th]]. 




\bibitem{SSB}
F. Benini, S. Benvenuti and Y. Tachikawa, ``Webs of five-branes and N=2 superconformal field theories'', JHEP 0909, 052 (2009), [arXiv:0906.0359 [hep-th]].







\bibitem{S}
F.\ Benini, Y.\ Tachikawa, and B.\ Wecht, ``Sicilian gauge theories and N=1 dualities'', JHEP 01 (2010) 088, [arXiv:0909.1327 [hep-th]].

\bibitem{T}
L. F. Alday, F. Benini and Y. Tachikawa, ``Liouville/Toda central charges from M5-branes'', Phys. Rev. Lett. 105, 141601, (2010), [arXiv:0909.4776 [hep-th]].



\bibitem{U}
Y. T. Huang and A. E. Lipstein, ``Amplitudes of 3D and 6D Maximal Superconformal Theories in Supertwistor Space'', JHEP 1010, 007, (2010), [arXiv:1004.4735 [hep-th]].

\bibitem{X1}
J. Bagger and N. Lambert, ``Modeling multiple M2's'', Phys. Rev. D 75, 045020 (2007), hep-th/0611108. 


\bibitem{X2}
J. Bagger and N. Lambert, ``Gauge Symmetry and Supersymmetry of Multiple M2-Branes'', Phys. Rev. D 77, 065008 (2008) [arXiv:0711.0955 [hep-th]]. 

\bibitem{X3}
A. Gustavsson, ``Algebraic structures on parallel M2-branes'', Nucl. Phys. B 811, 66 (2009) [arXiv:0709.1260 [hep-th]]. 


\bibitem{Y}
O. Aharony, O. Bergman, D. L. Jafferis and J. Maldacena, ``N = 6 superconformal Chern-Simons-matter theories, M2-branes and their gravity duals'', JHEP 10, 091, (2008), [arXiv:0806.1218 [hep-th]]. 



\bibitem{a}
C. Papageorgakis and C. Saemann, ``The 3-Lie algebra (2,0) tensor multiplet and equations of motion on loop space'', JHEP 1105, 099, (2011), [arXiv:1103.6192 [hep-th]].








\bibitem{V}
N. Lambert, D. Tong, ``Membranes on an Orbifold'', Phys. Rev. Lett. 101, 041602 (2008), [arXiv:0804.1114 [hep-th]].


\bibitem{W}
D. Berenstein and D. Trancanelli, ``Three-dimensional N=6 SCFT's and their membrane dynamics'', Phys. Rev. D78, 106009 (2008), [arXiv:0808.2503 [hep-th]].








\bibitem{b}
K.-W. Huang and W.-H. Huang, ``Lie 3-algebra non-abelian (2,0) theory in loop space'', [arXiv:1008.3834 [hep-th]].

\bibitem{cccc}
H. Singh, ``Super-Yang-Mills and M5-branes'', JHEP 08 (2011) 136, [arXiv:1107.3408 [hep-th]].



\bibitem{ccc}
N. Lambert, C. Papageorgakis, and M. Schmidt-Sommerfeld, ``M5-Branes, D4-Branes and Quantum 5D Super-Yang-Mills'', JHEP 01 (2011) 083, [arXiv:1012.2882 [hep-th]].


\bibitem{Ber}
D. S. Berman and J. A. Harvey, ``The self-dual string and anomalies in the M5-brane'', JHEP 0411, 015 (2004), hep-th/0408198.




\bibitem{B}
M.\ J. Duff and J.\ X. Lu, ``Black and super p-branes in diverse dimensions'', \NP {\bf B416} 301 (1994), hep-th/9306052.











\end{thebibliography}
\end{document}